\documentclass[preprint,times,tighten]{aastex63}

\usepackage{epsf,color,amsmath}
\usepackage{multirow}
\usepackage{cancel}

\newcommand{\sfrac}[2]{\mathchoice%
  {\kern0em\raise.5ex\hbox{\the\scriptfont0 #1}\kern-.15em/
    \kern-.15em\lower.25ex\hbox{\the\scriptfont0 #2}}
  {\kern0em\raise.5ex\hbox{\the\scriptfont0 #1}\kern-.15em/
    \kern-.15em\lower.25ex\hbox{\the\scriptfont0 #2}}
  {\kern0em\raise.5ex\hbox{\the\scriptscriptfont0 #1}\kern-.2em/
    \kern-.15em\lower.25ex\hbox{\the\scriptscriptfont0 #2}} {#1\!/#2}}

\newcommand{\Ub}{{\bf{U}}}

\newcommand{\Omegab}{{\bf{\Omega}}}
\newcommand{\gb}{{\bf{g}}}
\newcommand{\rb}{{\bf{r}}}

\newcommand{\epsdot}{\dot{\epsilon}}

\newcommand{\nablab}{\mathbf{\nabla}}

\newcommand{\omegadot}{\dot{\omega}}

\newcommand{\kth}{k_{\rm th}}

\newcommand{\C}{$^{12}$C}

\newcommand{\isot}[2]{$^{#2}\mathrm{#1}$}
\newcommand{\isotm}[2]{{}^{#2}\mathrm{#1}}

\newcommand{\castro}{{\sf Castro}}
\newcommand{\amrex}{{\sf AMReX}}
\newcommand{\pynucastro}{{\sf pynucastro}}

\newcommand{\gcc}{\mathrm{g~cm^{-3} }}
\newcommand{\cms}{\mathrm{cm~s^{-1} }}

\begin{document}
\title{Dynamics of Laterally Propagating Flames in X-ray Bursts. I. Burning Front Structure}

\shorttitle{Lateral Flame Dynamics}
\shortauthors{Eiden et al.}

\author[0000-0001-6191-4285]{Kiran Eiden}
\affiliation{Dept.\ of Physics and Astronomy, Stony Brook University,
             Stony Brook, NY 11794-3800}

\author[0000-0001-8401-030X]{Michael Zingale}
\affiliation{Dept.\ of Physics and Astronomy, Stony Brook University,
             Stony Brook, NY 11794-3800}
\affiliation{Center for Computational Astrophysics, Flatiron Institute, New York, NY 10010}

\author[0000-0002-1530-781X]{Alice Harpole}
\affiliation{Dept.\ of Physics and Astronomy, Stony Brook University,
             Stony Brook, NY 11794-3800}

\author[0000-0003-2300-5165]{Donald Willcox}
\affiliation{Lawrence Berkeley National Laboratory, Berkeley, CA}

\author[0000-0002-6447-3603]{Yuri Cavecchi}
\affiliation{Mathematical Sciences and STAG Research Centre, University of Southampton, SO17 1BJ}

\author[0000-0003-0439-4556]{Max P.\ Katz}
\affiliation{NVIDIA Corp}

\correspondingauthor{Kiran Eiden}
\email{kiran.eiden@stonybrook.edu}

\begin{abstract}
We investigate the structure of laterally-propagating flames through
the highly-stratified burning layer in an X-ray burst.
Two-dimensional hydrodynamics simulations of flame propagation are
performed through a rotating plane-parallel atmosphere, exploring the
structure of the flame.  We discuss the approximations needed to
capture the length and time scales at play in an X-ray burst and
describe the flame acceleration observed.  Our studies complement
other multidimensional studies of burning in X-ray bursts.
\end{abstract}


\section{Introduction}\label{Sec:Introduction}

X-ray bursts (XRBs) are thermonuclear explosions in an accreted H or
He layer on the surface of a neutron star (see~\citealt{galloway:2017}
for a review).  Observations of bursts can assist in constraining the
properties of the underlying neutron star, helping to illuminate the
nuclear equation of state~\citep{steiner:2010,ozel2016masses}.  Extensive
observations of brightness oscillations during the rise (the initial phase
of the burst where the observed flux rapidly increases) have provided
evidence that the burning begins in a localized region and spreads
over the surface of the neutron
star~\citep{bhattacharyya:2006,bhattacharyya:2007,chakraborty:2014}.

One dimensional studies of XRBs have been very successful in predicting the
lightcurves and recurrence times (see, e.g., \citealt{woosley-xrb}).
These assume spherical symmetry and thus cannot capture the effects of
localized burning spreading across the neutron star.  These studies
have also been used to explore the sensitivity of the burst
observables to accretion and reaction
rates~\citep{cyburt:2010,Jose2010a,Lampe2016}, and to model individual
bursts~\citep{johnston:2019}.

Multidimensional simulations of burning on a neutron star are more
difficult, with both the temporal and spatial scales presenting
challenges (see \citealt{astronum_2018} for an overview).  For the spatial scales, we need to
resolve the reaction zone, $\mathcal{O}(10~\mbox{cm})$ or smaller, the
scale height of the atmosphere, $\mathcal{O}(500~\mbox{cm})$, and the
Rossby scale where the Coriolis force balances the lateral pressure
gradient, $\mathcal{O}(10^5~\mbox{cm})$~\citep{spitkovsky2002}.  For
the temporal scales, capturing the rise, $\mathcal{O}(1~\mbox{s})$,
and the decay of the burst, $\mathcal{O}(10~\mbox{s})$, as well as the
accretion period between bursts, $\mathcal{O}(10^4~\mbox{s})$, is
currently beyond the ability of multidimensional hydro codes.
Nevertheless, significant progress has been made in understanding the
multidimensional nature of XRBs, through various approximations.

Laterally propagating detonations were modeled by
\citet{fryxellwoosley82} and \citet{hedet}. However, since it is difficult to detonate helium at the
densities found in normal XRBs, and even harder to detonate hydrogen
because of the waiting times for weak reactions, these would only occur at
very high densities.  This means that
detonations may only really be applicable to superbursts (where carbon
is the reactant)~\citep{Weinberg2006b,Weinberg2007}.

Global multidimensional studies were performed by
\citet{spitkovsky2002}, where it was demonstrated that the Coriolis
force plays an important role in confining the burning as it spreads
across the neutron star surface.  These calculations used the
shallow-water approximation, so the vertical details of the
atmosphere's structure were not captured.  Their model showed that the
horizontal pressure gradient between the ash and fuel can be important
in accelerating the burning front.

Small-domain studies of convective burning in XRBs preceding flame
development have been done in two-dimensions \citep{lin:2006,xrb,xrb2}
and three-dimensions \citep{xrb3d}.  These calculations used low Mach
number methods, which approximate the hydrodynamics equations to filter
soundwaves, enabling large timesteps and efficient modeling of
subsonic convection.  While these calculations could not support
the lateral differences needed for flame spreading, they can help
understand the role that convection plays in distributing the initial
burning products vertically throughout the neutron star atmosphere as well
as the nature of any turbulence the burning front might encounter as it propagates
through the atmosphere.

The first vertically resolved simulations of lateral deflagrations
were obtained by \citet{cavecchi:2013}, who showed how the Coriolis
force creates a geometrical configuration that increases the flame
speed set by conduction by a factor $\sim L_R / H$, where $L_R$ is the
Rossby radius and $H$ is the scale height of the burning layer. The
effect of changing Coriolis confinement across the surface on the flame 
propagation was explored in \citet{art-2015-cavecchi-etal},
while \citet{art-2016-cavecchi-etal} showed how magnetic field tension,
opposing the Coriolis force, can either speed up or slow down the flame by
changing the horizontal extent of the flame front. Finally, \citet{Cavecchi2019}
studied the effects in 3D of the baroclinic instability at the flame
front, measuring flames up to 10 times faster than in the 2D case.

For deflagrations, we either need to resolve the structure of the
reaction zone or use a flame model.  Flame models usually assume that
the flame structure is thin compared to the size of the system (see,
e.g., \citet{Ropke2007} for applications to Type Ia supernovae).  For
XRBs however, the flame thickness is comparable to the scale height of
the atmosphere, so we cannot use these approximations.  Accurate
models of flames in XRBs therefore require that we resolve the thermal width,
which is $\mathcal{O}(10~\mbox{cm})$ for helium flames~\citep{Timmes00}.

The goal of this study is to understand what numerical and physical
approximations are required to perform a full hydrodynamical,
multidimensional simulation of flame propagation through the
atmosphere of a neutron star.  For simplicity in this first set of
calculations, we will use a pure helium composition.  These studies
complement the prior multidimensional studies described above in helping
us to build a picture of the dynamics of X-ray bursts.

\section{Numerical Approach}\label{Sec:numerics}

All simulations are performed with the \castro\ hydrodynamics
code~(\citealt{castro}; see also \citealt{astronum:2017} for a recent
description).  We evolve the system of fully compressible Euler
equations for reacting flow:
\begin{eqnarray}
\frac{\partial( \rho X_k)}{\partial t} &=& - \nabla\cdot (\rho \Ub X_k) + \rho \omegadot_k \\
\frac{\partial (\rho \Ub)}{\partial t} &=& - \nabla\cdot (\rho \Ub \Ub) - \nabla p +
    \rho \gb \nonumber \\
  &&-2 \rho \Omegab\times \Ub - \rho \Omegab \times (\Omegab \times \rb) \\
\frac{\partial (\rho E)}{\partial t} &=& - \nabla \cdot (\rho \Ub E + p\Ub) +
    \nabla \cdot \kth \nabla T + \rho \epsdot +\nonumber\\
  && \rho \Ub \cdot \gb - \rho (\Omegab \cdot \rb)(\Omegab \cdot \Ub) + \rho |\Omegab|^2 (\Ub \cdot \rb)
\end{eqnarray}
Here, $\rho$ is the mass density, $\Ub$ is the velocity, $p$ is the
pressure and $E$ is the specific total energy, which is related to the
specific internal energy as $e = E - |\Ub|^2/2$.  The forcing in the
momentum equation includes gravity, described by gravitational
acceleration $\gb$, and rotational forces, described by angular velocity
$\Omegab$, with $\rb$ the position vector from the origin.
Species are described by mass fractions, $X_k$ (such that $\sum_k X_k
= 1$), and creation rates, $\omegadot$, and are related to the total
specific energy generation rate, $\epsdot$.  The total mass
conservation,
\begin{equation}
\frac{\partial \rho}{\partial t} = - \nabla\cdot (\rho \Ub)
\end{equation}
implies $\sum_k \omegadot_k = 0$.  An equation of state of the form $p = p(\rho, e, X_k)$ completes the
thermodynamic description of the system.  Thermal diffusion is 
described by a thermal conductivity $\kth$ and temperature $T$. 

\castro\ uses an unsplit piecewise
parabolic method (PPM) with characteristic tracing for solving the
hydrodynamics~\citep{ppm,millercolella:2002}, generalized to an
arbitrary equation of state~\citep{zingalekatz}.  Reactions are
incorporated via Strang splitting~\citep{strang:1968}, giving a method
that is overall second-order accurate in space and time.
\castro\ uses the \amrex\ adaptive mesh refinement
library~\citep{amrex_joss} to manage a hierarchy of grids at different
resolutions.

Since the neutron star rotates, we work in a corotating frame, taking
the angular velocity $\Omegab$ to be constant.  Further, for the two-dimensional
simulations presented here, we work in axisymmetric coordinates,
but we advect a third component of velocity, coming out of the
simulation plane, that participates in the Coriolis force (sometimes
described as a 2.5D simulation).  We will take the
\castro\ $x$-coordinate to be the cylindrical radial coordinate with
corresponding velocity $u$, the \castro\ $y$-coordinate to be the
cylindrical vertical coordinate with corresponding velocity $v$, and the
\castro\ $z$-coordinate to be the cylindrical azimuthal coordinate,
with corresponding velocity $w$.  A righthanded coordinate system has
positive $w$ pointing out of the page.  We will take $\Omegab =
\Omega_0 \hat{\bf y}$ for the angular rotation rate, and $\gb = -g
\hat{\bf y}$ for the gravitational acceleration, with $g$ constant.
With these choices, the Coriolis force is:
\begin{equation}
-2\rho \Omegab \times \Ub =
   -2\rho \left ( \Omega_0 w \hat{x} - \Omega_0 u \hat{z} \right )
\end{equation}
We will neglect the centrifugal force---with our plane-parallel
geometry, this will act only in the lateral direction, and is not
expected to greatly affect the dynamics.  Carrying the Coriolis force
allows us to capture the geostrophic balance that sets up via lateral
hydrostatic equilibrium \citep{spitkovsky2002}.  In the discussions
below, we will use the Castro coordinate names, $(x, y)$ in our
notation.

Writing the momentum equation in terms of the $u$, $v$, and $w$
components, and neglecting the centrifugal force, we have:
\begin{align}
\frac{\partial (\rho u)}{\partial t} + \nabla \cdot (\rho u \Ub) +
     \frac{\partial p}{\partial x} &= -2\rho \Omega_0 w  \\
\frac{\partial (\rho v)}{\partial t} + \nabla \cdot (\rho v \Ub) +
     \frac{\partial p}{\partial y} &= -\rho g \\
\frac{\partial (\rho w)}{\partial t} + \nabla \cdot (\rho w \Ub) +
     \cancelto{0}{\frac{\partial p}{\partial z}} &=
    2\rho \Omega_0 u
\end{align}
where we cancel $\partial p/\partial z$ because there are no
variations in the azimuthal direction.  This allows us to recast the
$w$-velocity equation as a simple advection equation:
\begin{equation}
\frac{\partial w}{\partial t} + \Ub \cdot \nabla w = 2\Omega_0 u
\end{equation}
In our geometry, the flame will propagate from left to right, so $u$
will be positive and the Coriolis force results in $w > 0$ (out of the
simulation plane).  The algorithmic implementation of rotation in
\castro\ is described in \cite{wdmergerI}.

We use a general stellar equation of state with nuclei (treated as an
ideal gas), photons, and degenerate/relativistic electrons, as
described in \cite{timmes_swesty:2000}.  To model reactions, we use a 
13-isotope alpha chain network derived from the {\tt aprox13}
network~\citep{timmes_aprox13}.  For one of our runs, we use the
smaller 7-isotope network described in \citet{iso7}.  We integrate the
network using the VODE integration package~\citep{vode}, and our
implementation is provided in the StarKiller Microphysics
source~\citep{starkiller}.
We note that we do not explicitly model viscosity.  The reactions will
provide the smallscale cutoff to the instabilities and turbulence at
the flame front.  We also do not include species
diffusion---astrophysical flames tend to have large Lewis numbers, so
this is not expected to be important~\citep{timmeswoosley:1992}.
Finally, we use the thermal conductivities described in
\citet{Timmes00}.

\begin{figure}[t]
	\plotone{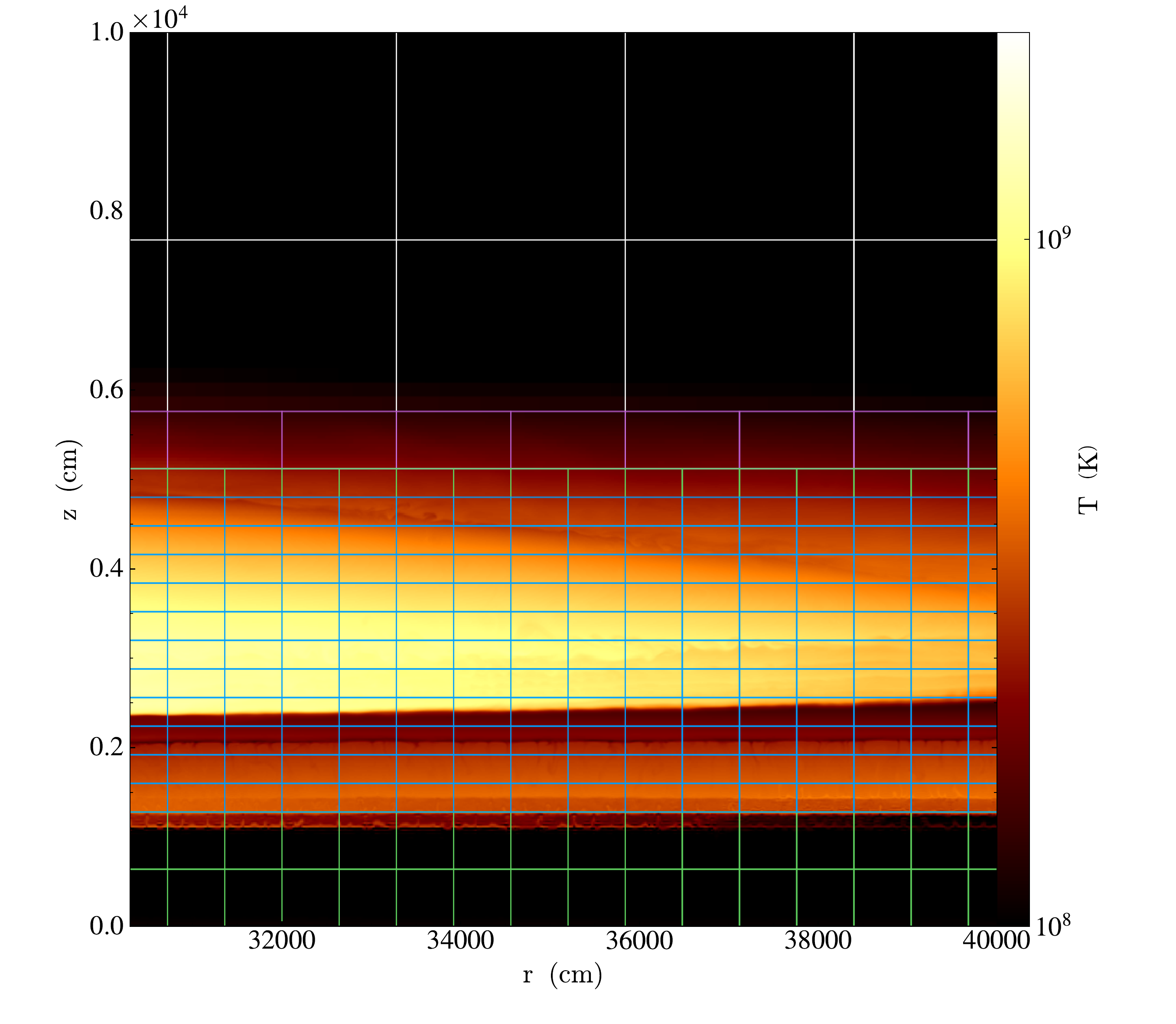}
	\caption{\label{fig:grids}A section of a 2D simulation showing the four-level
		grid structure. Note that the boxes shown are not the simulation zones, which span
		$10~\mathrm{cm}$ at the finest level, but subdomains containing approximately equal
        numbers of zones that are distributed across MPI processes. A coarse base grid
		extending to the upper boundary is drawn in white, with the magenta, green, and
		blue grids showing the jumps in refinement needed to fully resolve the fuel layer
		and underlying neutron star.}
\end{figure}

All simulations use adaptive mesh refinement to refine on the
atmosphere (leaving the space between the top of the atmosphere and
upper boundary at low resolution). As seen in Figure \ref{fig:grids}, 
we use up to 3 refinement levels in addition to the base grid, the first
one a factor of 4 finer than the previous and the remaining each a
factor of 2 finer than the previous. \castro\ subcycles in time, so the 
finer grids are evolved with a finer timestep than the coarse grids. 
Occasionally, the timestep chosen at the start of a cycle will violate
the CFL condition during the advancement of the finer grids. In this
case, we restart the finer grid evolution with a smaller timestep,
subcycling within the larger timestep hierarchy.  We use a CFL number
of 0.8 for our simulations.  
The base grid for our standard
simulations is $768\times 192$ zones, and the equivalent finest grid
when 3 refinement levels are added would be $12288\times 3072$ zones.
Our standard domain is $1.2288\times 10^5~\mathrm{cm} \times
3.072\times 10^4~\mathrm{cm}$, corresponding to $10~\mathrm{cm}$ resolution
on the finest grid.  We only refine the fuel layer in the left half of
the domain at the highest resolution (and only down to densities of
$2.5\times 10^4~\gcc$), since this is where we expect the flame to
propagate.  At the start of the simulation, $3.4\%$ of the domain is
at the finest resolution.  This increases to $7.4\%$ by the end of the
simulation, because of the increase in the scale height of the atmosphere
behind the flame.

Thermal diffusion is modeled explicitly, using a predictor-corrector
scheme to achieve second-order accuracy.  A verification test of the
diffusion scheme is shown in Appendix~\ref{app:diffusion}.  The
explicit thermal diffusion requires a timestep limiter of the form:
\begin{equation}
\Delta t_\mathrm{diff} \le \frac{1}{2} \frac{\Delta x^2}{\mathcal{D}}
\end{equation}
where $\mathcal{D} = \kth/(\rho c_v)$ is the thermal diffusivity.  The
diffusivity increases rapidly at the top of the atmosphere, causing
these low density regions to determine the overall timestep for the
simulations.  Therefore, we disable thermal conduction at low
densities where it is not expected to be important.

We use hydrostatic boundary conditions on the lower boundary, using
a discretized hydrostatic equilibrium equation of the form:
\begin{equation}
\label{eq:hse}
    p_i = p_{i-1} + \frac{1}{2} \Delta y (\rho_i + \rho_{i-1}) \gb \cdot \hat{\bf y}
\end{equation}
and holding the temperature constant in the ghost cells.  This is solved
together with the equation of state.  The velocity is reflected at this boundary.
This procedure follows the form described in \citet{ppm-hse}.  The
left boundary is reflecting and the right boundary is a zero-gradient
outflow.  The top boundary sets the state to simply the conditions
in our outer buffer region of the initial model (see below), with the
normal velocity set to the larger of zero or the velocity at the top
of the domain (this prevents incoming velocities at the top) and the
transverse velocities set to zero.

When we begin the simulation, there is a transient phase as the flame
gets established.  Material that is forced upward will encounter the
steep density gradient at the top of the atmosphere and accelerate as
it is blown out of the atmosphere.  Eventually this material will fall
back to the top of the atmosphere.  In this paper, we are mostly
concerned with the behavior of the flame and not any material that is
violently blown out of the top of the atmosphere, so we apply a sponge
to this region.  This is similar to the method we previously used in \cite{xrb2}, and takes
the form of a source term to the momentum and energy equations of the form:
\begin{align}
{\bf S}_{\rho \Ub} &= \rho \Ub \frac{f}{\Delta t} \\
S_{\rho E} &= \rho \Ub \cdot {\bf S}_{\rho \Ub}
\end{align}
with the sponge forcing $f$ dependent on the density.  We define the sponge
shape as:
\begin{equation}
s = \left \{
     \begin{array}{cc}
            0   & \rho > \rho_\mathrm{upper} \\
            \frac{1}{2}
                \left [ 1 - \cos \left ( \frac{\pi (\rho - \rho_\mathrm{upper})}{\Delta \rho} \right ) \right ]  & \rho_\mathrm{upper} \ge \rho > \rho_\mathrm{lower} \\
                     1 & \rho < \rho_\mathrm{lower} 
             \end{array} \right .
\end{equation}
Here $\rho_\mathrm{upper}$ and $\rho_\mathrm{lower}$ are the densities where the sponge transitions to being fully applied.  We take $\rho_\mathrm{upper} = 10^2~\gcc$ and $\rho_\mathrm{lower} = 1~\gcc$, with $\Delta \rho = \rho_\mathrm{upper} - \rho_\mathrm{lower}$.
The sponge update is done implicitly to get the effective forcing, $f$:
\begin{equation}
f = -\left [ 1 - \frac{1}{1 + \alpha s} \right ]
\end{equation}
with $\alpha = \Delta t/\tau_\mathrm{sponge}$.  Here
$\tau_\mathrm{sponge}$ is the timescale over which the sponge acts.
We take $\tau_\mathrm{sponge} = 10^{-7}~\mathrm{s}$.
The sponge drives the velocity of the material in the low
density regions at the top
of the atmosphere to zero.  This sponging helps increase our timestep
as well.

\section{Flame Properties}\label{Sec:Flame}

The speed and thickness of a laminar helium flame are determined by the
energy generation rate and conductivity, and scale roughly as
\begin{equation}
\label{eq:flame_scaling}
s_L \approx \sqrt{\kth \epsdot} \qquad
\lambda_L \approx \sqrt{\frac{\kth}{\epsdot}}
\end{equation}
\citep{orourke:1979,khokhlov:1993}.
At the densities we consider in this simulation, the pure He flame
speed is quite slow and would require long integration
times to see significant evolution of the burning.  We therefore
consider boosted flames in this first paper, to accelerate the burning
and allow us to understand the qualitative effects of laterally
propagating flames.  To boost the flame while keeping the thickness
the same, we can multiply both the burning rate and the conductivity
by the same factor.  For our standard calculations, we choose 10 for
each, to give a 10$\times$ faster flame speed.  We call this the ``10/10''
flame.  We will also do a simulation with the reactions and
conductivity both boosted by 5, the ``5/5'' flame.

\begin{figure*}[t]
\plottwo{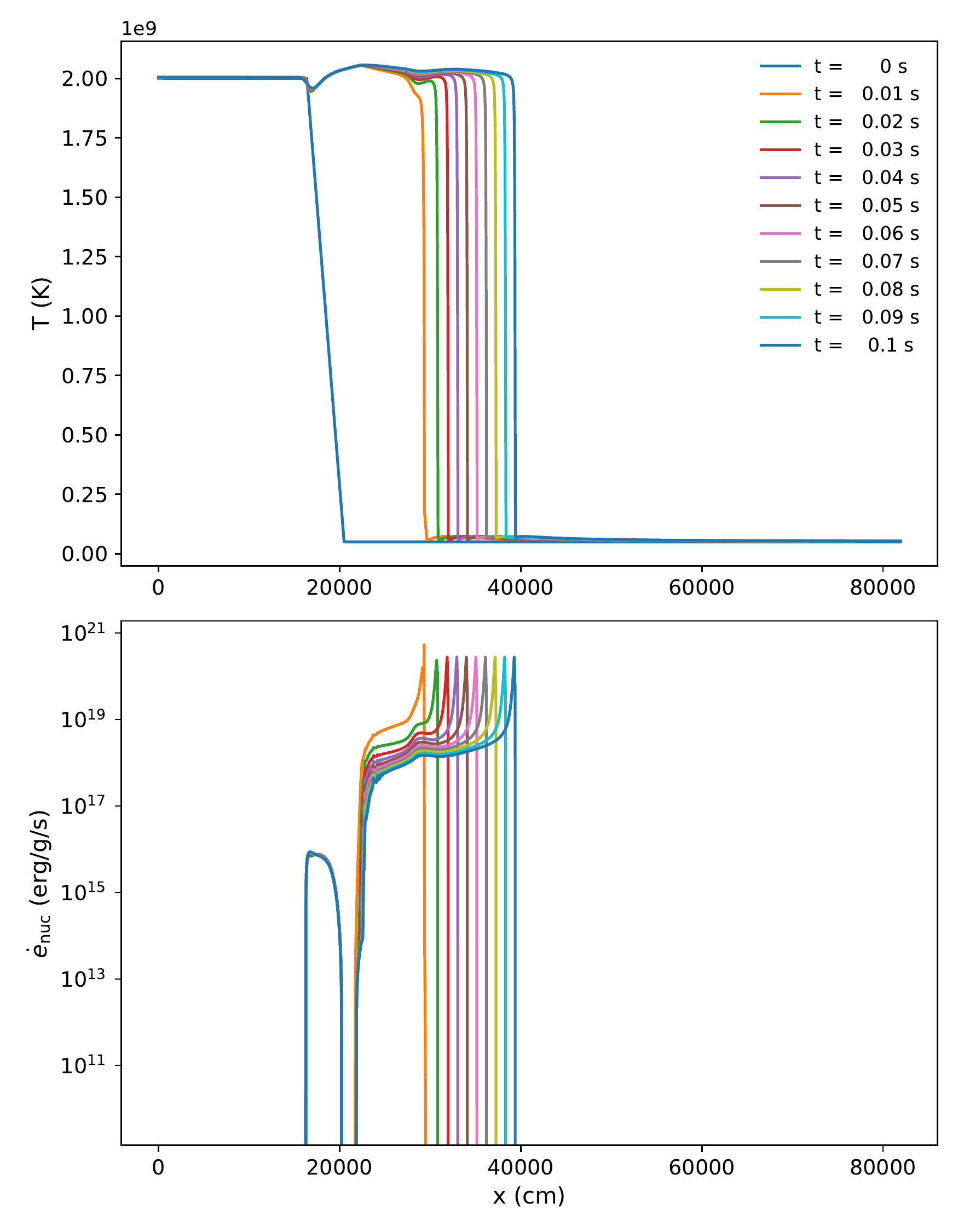}{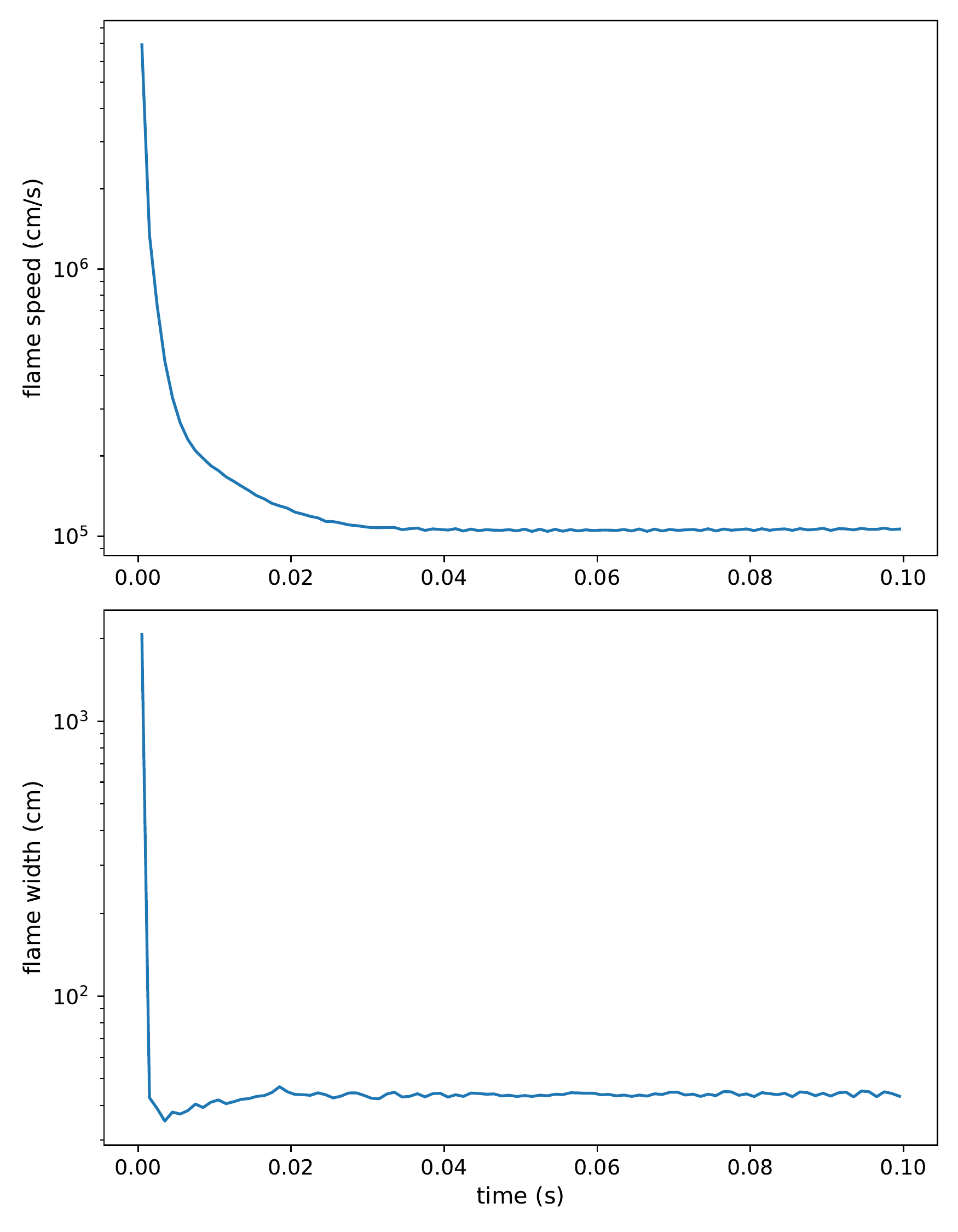}
\caption{\label{fig:flame} Time-evolution of the 10$\times$ boosted 1D
	laminar flame. The left plot shows temperature and nuclear energy
	generation profiles at 11 different times, while the right plot
	shows flame propagation speed and flame thickness as functions of time.}
\end{figure*}

To understand the time and length scales involved in flame
propagation, we do a 1D simulation of a laminar flame using our
microphysics.  Figure~\ref{fig:flame} shows the flame thermodynamic
profile and properties for the 10/10 flame using
our conductivities and the {\tt aprox13} reaction network.  This flame
had a density of $2\times 10^6~\gcc$ and
temperature of $5\times 10^7$~K.  We observe that this flame speed is
about $10^5~\cms$, the flame width is about 40~cm, and it
takes about 3~ms to settle into a sustained flame.  Note that this speed
is quite small compared to the speeds of $\sim 10^6~\cms$ estimated in
\citet{spitkovsky2002}.  Table~\ref{table:flame_speeds_1d} gives the properties
of the 10/10 and 5/5 laminar flames.  We expect a multidimensional flame to
accelerate due to hydrodynamics interactions (wrinkling, turbulence
interactions, directed flows feeding fuel into the flame, etc.).

We measure the laminar flame width as:
\begin{equation}
\lambda_L \equiv \frac{\Delta T}{\max\{|\nabla T|\}}
\end{equation}
Experience with modeling resolved flames suggests that we need a
spatial resolution, $\Delta x$ of $\lambda_L/\Delta x \sim 5$~\citep{SNld}.  These
conditions represent the bottom of the He layer.  As the density
decreases with altitude, the flame thickness increases and the flame
speed decreases, so we will easily resolve the flame structure throughout
the rest of the atmosphere.

\begin{deluxetable}{lc}
	\tablecaption{\label{table:flame_speeds_1d} Laminar flame speeds.}
	\tablehead{\colhead{run} & \colhead{$s_L$ (km s$^{-1}$)}}
	\startdata
	10/10 boost & $1.06 \pm 0.01$ \\
	5/5 boost & $0.56 \pm 0.01$ \\
	\enddata
\end{deluxetable}

\section{Initial Model}\label{Sec:inital_model}


We wish to create an initial atmosphere consisting of a hot
``post-flame'' region and a cooler atmosphere that the flame will
laterally propagate into.  We put the hot region at the very left of
the domain (the origin of the axisymmetric coordinates).  To create
these initial conditions, we produce two different hydrostatic models,
a ``hot'' model that will represent the perturbation that drives the
flame and a ``cool'' model that will represent the state ahead of the
flame.  These will have different scale heights.  To create these
models, we break the vertical structure of the atmosphere into four
layers: (1) the underlying neutron star, (2) a ramp-up to the base of the
accreted atmosphere, (3) a fuel layer representing the bulk of the
atmosphere where the flame will propagate, and (4) an outer, low density,
isothermal buffer above the atmosphere that allows us
to capture expansion and explosive dynamics.  

The temperature profile in the star and ramp region is given as:
\begin{equation}
T(y) = T_\star + \frac{1}{2} (T_\mathrm{hi} - T_\star) \left [ 1 + \tanh\left( \frac{\tilde{y}}{2 \delta_\mathrm{atm}} \right ) \right ]
\end{equation}
with
\begin{equation}
\tilde{y} = y - H_\star - \frac{3}{2} \delta_\mathrm{atm}
\end{equation}
Here, $\delta_\mathrm{atm}$ is a characteristic width of the
transition ramp, $T_\star$ is the temperature of the neutron star, and
$T_\mathrm{hi}$ is the highest temperature in the HSE model---it will
represent the base of the fuel layer.

The species mass fractions use this same profile, switching from a set
describing the underlying star, ${X_k}_\star$, and the set for the
accreted material, ${X_k}_\mathrm{atm}$, which is used in the
isentropic and outer regions.  Note, since the profile above is
linear in $X$, if the initial mass fractions sum to one, then the blended
mass fractions in the ramp region also sum to one.

We specify the density, $\rho_\mathrm{int} = \rho(y = H_\star)$ as the
starting point for the integration of hydrostatic equilibrium.  This
is just below the ramp-up region---this ensures that regardless of
what the peak temperature ($T_\mathrm{hi}$) is, the state beneath the
ramp-up region remains unchanged.  Therefore, we will still be in
lateral equilibrium in the star region.  We will denote the density
where $T = T_\mathrm{hi}$ as $\rho_\mathrm{fuel}$.

\begin{figure}[t]
\centering
\epsscale{0.75}
\plotone{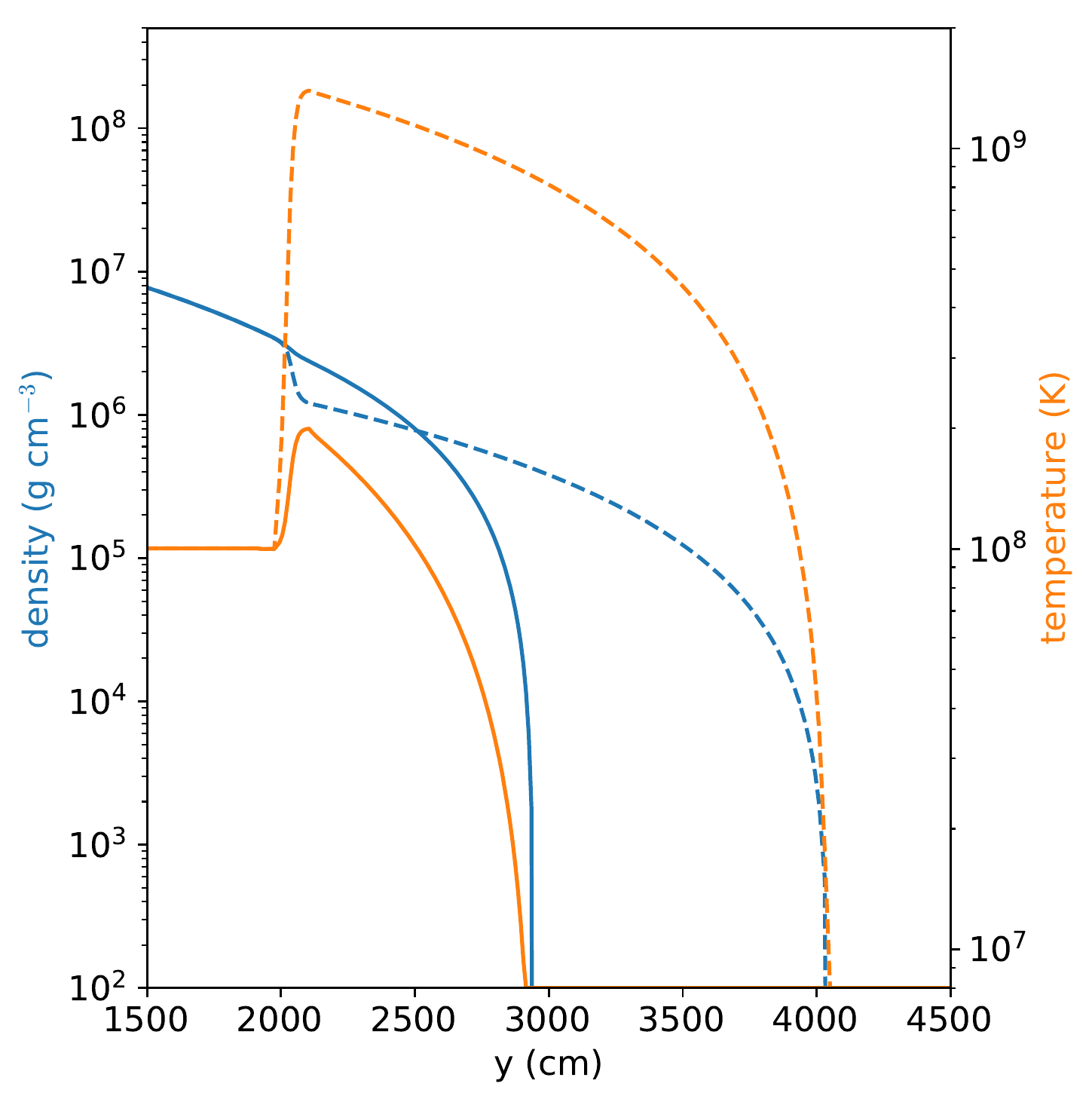}
\caption{\label{fig:initial_models} Our ``cool'' (solid) and ``hot''
  (dashed) initial models, showing both the density and temperature.}
\epsscale{1.0}
\end{figure}

Creating the model involves specifying $T_\star$, $T_\mathrm{hi}$,
$T_\mathrm{lo}$, $\rho_\mathrm{int}$, $H_\star$,
$\delta_\mathrm{atm}$, ${X_k}_\star$, ${X_k}_\mathrm{atm}$, and $g$.  We
then integrate outwards from the base of the ramp region ($y =
H_\star$), enforcing the discrete form of hydrostatic equilibrium,
Eq.~\ref{eq:hse}.  Integrating upwards, we find $p_i$ and
$\rho_i$ using a Newton-Raphson solver together with the equation of state, with either the temperature
specified, $T_i = T(p_i, \rho_i, \{X_k\}_i)$ (in the isothermal, ramp,
and buffer layers) or constant entropy, $s_i = s(p_i, \rho_i,
\{X_k\}_i)$, in the fuel layer.  This follows the procedures described in \citet{ppm-hse}.  We
use a constant temperature for all $y < H_\star +
3\delta_\mathrm{atm}$.  Above this, we switch to an isentropic atmosphere until the
temperature drops to a floor value, $T_\mathrm{lo}$, at which point we
again keep the temperature constant.  The integration of the
atmosphere continues until the density falls to a low density cutoff,
$\rho_\mathrm{cutoff}$.  The material above this height is taken to
have constant density and temperature.
The choice of factors in front of $\delta_\mathrm{atm}$ were designed
to make sure the peak $T$ is attained at the desired density
of the burning layer.
The parameters we use for
the model generation are listed in Table~\ref{table:params} and the 
initial model profiles are showing in Figure~\ref{fig:initial_models}.

\begin{deluxetable}{lcc}
\tablecaption{\label{table:params} Initial model parameters.}
\tablehead{\colhead{parameter} & \colhead{cool} & \colhead{hot}}
\startdata
$T_\star$       & \multicolumn{2}{c}{$10^8$~K} \\
$T_\mathrm{hi}$ & $2\times 10^8$~K & $1.4\times 10^9$~K \\
$T_\mathrm{lo}$ & \multicolumn{2}{c}{$8\times 10^6$~K} \\
$\rho_\mathrm{int}$ & \multicolumn{2}{c}{$3.43\times 10^6~\gcc$} \\
$\rho_\mathrm{fuel}$\tablenotemark{a} & $2.36\times 10^6~\gcc$ & $1.20\times 10^6~\gcc$ \\
$\rho_\mathrm{cutoff}$ & \multicolumn{2}{c}{$10^{-4}~\gcc$} \\
$H_\star$       & \multicolumn{2}{c}{2000~cm} \\
$\delta_\mathrm{atm}$ & \multicolumn{2}{c}{50~cm} \\
$X_\star(\isotm{Ni}{56})\tablenotemark{b}$ & \multicolumn{2}{c}{1.0} \\
$X_\mathrm{atm}(\isotm{He}{4})\tablenotemark{b}$ & \multicolumn{2}{c}{1.0} \\
$\gb$             & \multicolumn{2}{c}{$-1.5\times 10^{14}~\mathrm{cm~s^{-2}} \hat{\bf y}$} \\
\enddata
\tablenotetext{a}{This is not an input parameter, but instead is
  computed during integration.  We list it here for reference.}
\tablenotetext{b}{All other species are taken as 0.}
\end{deluxetable}

We blend the hot and cold models laterally to produce the perturbation
needed to initiate a localized flame, with the hot model at the
origin of the axisymmetric geometry.  The blending is done as:
\begin{align}
p(x,y) &= f(x) p_\mathrm{hot}(y) + [1-f(x)] p_\mathrm{cool}(y) \\
\rho(x,y) &= f(x) \rho_\mathrm{hot}(y) + [1-f(x)] \rho_\mathrm{cool}(y) \\
X_k(x,y) &= f(x) {X_k}_\mathrm{hot}(y) + [1-f(x)] {X_k}_\mathrm{cool}(y)
\end{align}
with
\begin{equation}
f(x) = \begin{cases}
     1 & x < x_\mathrm{pert} \\
   1 - \frac{x - x_\mathrm{pert}}{\delta_\mathrm{blend}} & x_\mathrm{pert} \le x \le x_\mathrm{pert} + \delta_\mathrm{blend} \\
     0 & x > x_\mathrm{pert} + \delta_\mathrm{blend}
\end{cases}
\end{equation}
Since the equation of hydrostatic equilibrium is linear and our
blending is a linear combination of two models in hydrostatic
equilibrum, the blended model is also in vertical equilibrium initially.  We choose
$x_\mathrm{pert} = 1.024\times 10^4$~cm and $\delta_\mathrm{blend} = 2048$~cm.  Once
the blended model is constructed, we compute $T(x,y)$ and $(\rho e)(x,y)$
from the equation of state,
\begin{align}
  T(x,y) &= T(\rho(x,y), p(x,y), X_k(x,y)) \\
  (\rho e)(x,y) &= \rho(x, y) \cdot e(\rho(x,y), p(x,y), X_k(x,y)) 
\end{align}

The initial model is created on a uniform grid at the resolution
corresponding to the finest level of refinement.  At regions of the
atmosphere that are not refined to the finest level, we interpolate
density, pressure, and composition on the grid and then obtain the
temperature from the EOS.

\section{Simulations and Results}\label{Sec:results}

To assess the sensitivity of the flame propagation to the various approximations
we made, we ran a suite of simulations. 
Table~\ref{table:sim_names} summarizes these simulations.
The majority of them used a reaction rate boosting of $10$ and a
conductivity boosting of $10$, which should increase the flame speed
by a factor of $10$.  Most simulations used a resolution of
$10~\mbox{cm}$ and a domain width of a little more than one kilometer.  
We use an artifically high rotation rate of
$2000~\mbox{Hz}$, which gives a Rossby length of
\begin{equation}
L_R = \frac{\sqrt{g H_0}}{\Omega} \sim 3\times 10^4~\mathrm{cm}
\end{equation}
using a scale height $H_0 = 10^3~\mathrm{cm}$.  This is about one
quarter of the domain width.  The run at $20~\mbox{cm}$ resolution used
one fewer level of refinement.  The slower rotating case (1000~Hz)  uses a
slightly wider domain to accommodate the expected larger Rossby
length.  We note that the entire simulation framework for these
calculations is freely available in the \castro\ github
repository\footnote{\url{https://github.com/AMReX-Astro/Castro}, using
  the {\tt flame\_wave} setup.}.  In the discussions below, we'll use
the simulation names defined in the table to refer to specific runs
and we'll use the 10/10 run as the reference calculation.

\begin{deluxetable}{lcccccc}
\tablecaption{\label{table:sim_names} Simulation parameters.}
\tablehead{\colhead{name} &
           \colhead{reaction} &
           \colhead{conductivity} &
           \colhead{fine grid} &
           \colhead{rotation} &
           \colhead{domain size} &
           \colhead{network} \\
           \colhead{} &
           \colhead{boost} &
           \colhead{boost} &
           \colhead{resolution} &
           \colhead{rate} &
           \colhead{} &
           \colhead{}
}
\startdata
10/10 & 10 & 10 & 10~cm & 2000~Hz & $1.2288\times 10^5~\mbox{cm} \times 3.072\times 10^4~\mbox{cm}$ & {\tt aprox13} \\
5/5 & 5 & 5 & 10~cm & 2000~Hz & $1.2288\times 10^5~\mbox{cm} \times 3.072\times 10^4~\mbox{cm}$  &{\tt aprox13} \\
10/10-iso7 & 10 & 10 & 10~cm & 2000~Hz & $1.2288\times 10^5~\mbox{cm} \times 3.072\times 10^4~\mbox{cm}$ &{\tt iso7} \\
10/10-lowres & 10 & 10 & 20~cm & 2000~Hz & $1.2288\times 10^5~\mbox{cm} \times 3.072\times 10^4~\mbox{cm}$ &{\tt aprox13} \\
10/10-1000 Hz & 10 & 10 & 10~cm & 1000~Hz & $1.8432\times 10^5~\mbox{cm} \times 3.072\times 10^4~\mbox{cm}$ &{\tt aprox13} \\
\enddata
\end{deluxetable}

\subsection{General Features}

\begin{figure}[t]
\centering
\plotone{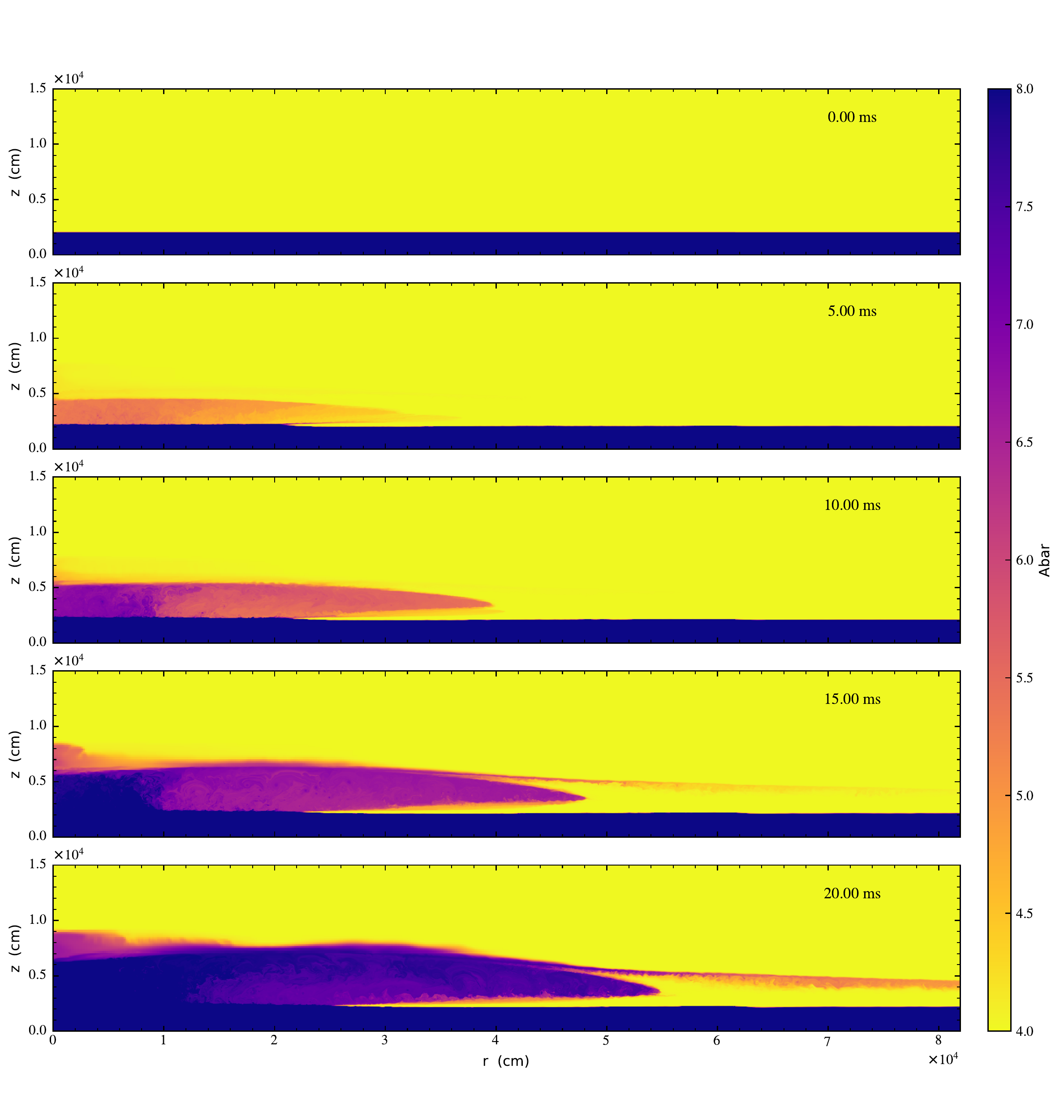}
\caption{\label{fig:time_series} Time series of the mean molecular weight of the 
flame for our standard 10/10 simulation.}
\end{figure}

Figure~\ref{fig:time_series} shows the time evolution of the 10/10
simulation, focusing on the mean molecular weight,
\begin{equation}
\bar{A} = \left ( \sum_k \frac{X_k}{A_k} \right )^{-1}.
\end{equation}
In each frame the buffer of \isot{Ni}{56} that serves as the
underlying neutron star is seen spanning the bottom of the domain.
Above that, the composition begins as pure \isot{He}{4}, but as the
simulation progresses, the flame processes this to heavier nuclei,
increasing $\bar{A}$.  By about $10~\mbox{ms}$, we see the flame front
is reasonably well-defined.  We see that the bottom of the burning
front is lifted off of the base of the atmosphere, greatly increasing
the surface area of the burning compared with a perfectly vertical
flame front.  By $20~\mbox{ms}$ the flame has moved out substantially
and we are beginning to see a gradient in the composition of the ash,
with the heavier nuclei furthest behind the flame.  The boosting of
the burning likely artificially increases this effect, as we'll see in the 5/5 case below.

\begin{figure}[t]
\centering
\plotone{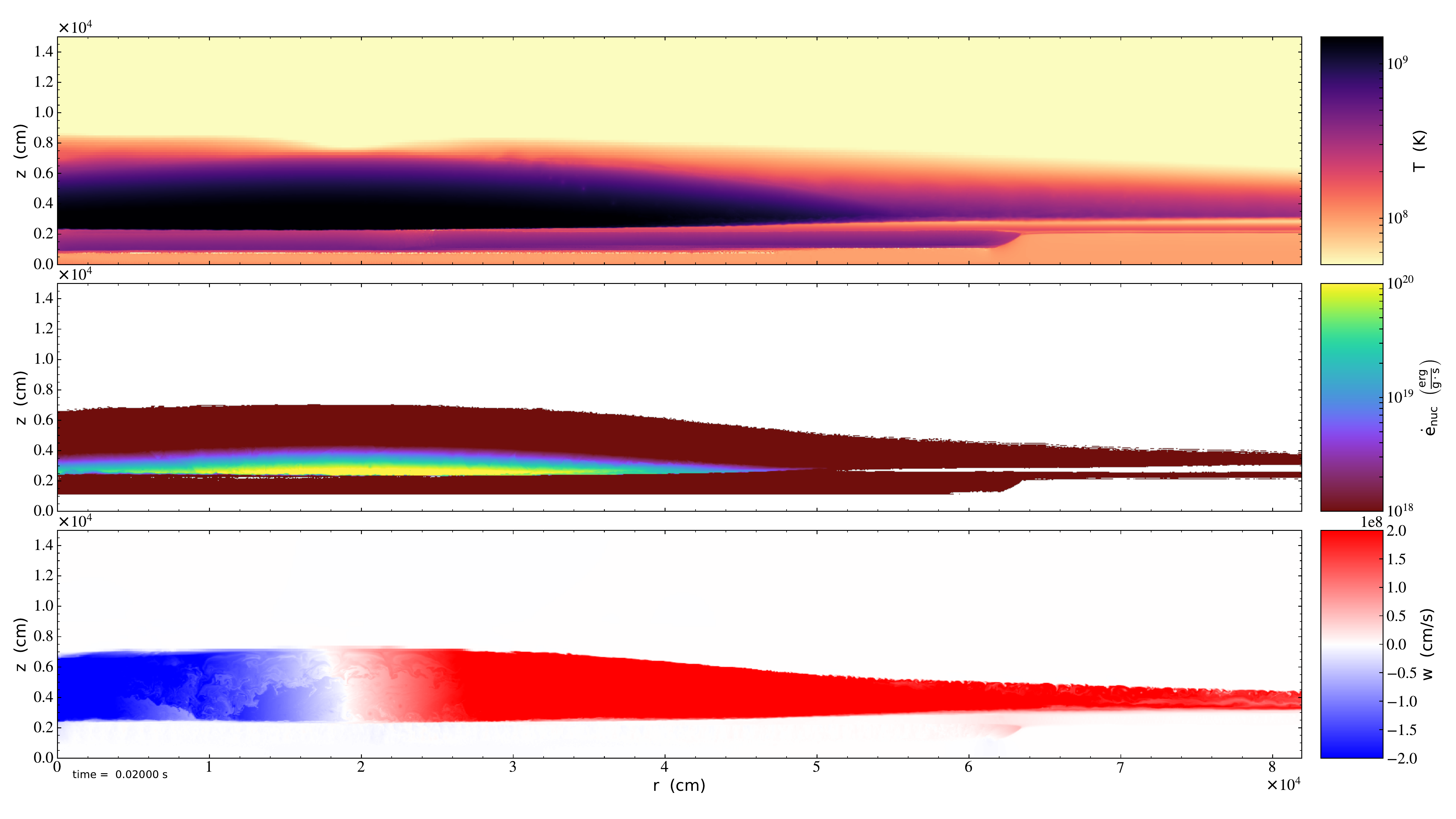}
\caption{\label{fig:10_10_overview} Temperature, energy generation rate, and out-of-plane velocity for the 10/10 simulation at 20~ms.}
\end{figure}

Figure~\ref{fig:10_10_overview} shows the temperature, energy
generation rate, and $w$ component of the velocity (the out-of-plane
velocity induced by the Coriolis force).  This latter field
illustrates the hurricane effect set up by the laterally spreading
burning front.  In the energy generation rate plot, we see that the
burning is mostly concentrated toward the bottom of the layer, as
expected since the density is greatest there.  We see that the peak of
the burning has moved off of the symmetry axis, demonstrating that the
burning front is propagating to the right.  In the temperature plot,
we see the effect of our refinement criteria focusing only on the part of
the atmosphere where we are most dense, with an artificial change in the
temperature at the refinement boundary due in part to the construction
of the initial model using the fine grid resolution for HSE.  As we will see in the lower resolution case, this does not
affect the results.

A final feature worth noting is the ash that seems to move along the
surface at a higher velocity than the flame, via surface gravity
waves.  The sponging that we perform is likely damping this to some extent, and
the method by which we initialize the flame may induce a larger
transient than in nature. Nevertheless, this surface ash is intriguing
because it affects the composition of the photosphere ahead of the
burning front, potentially changing our interpretation of
observations.  This is something that will be explored more fully in
the future.

\begin{figure}[t]
\centering
\plotone{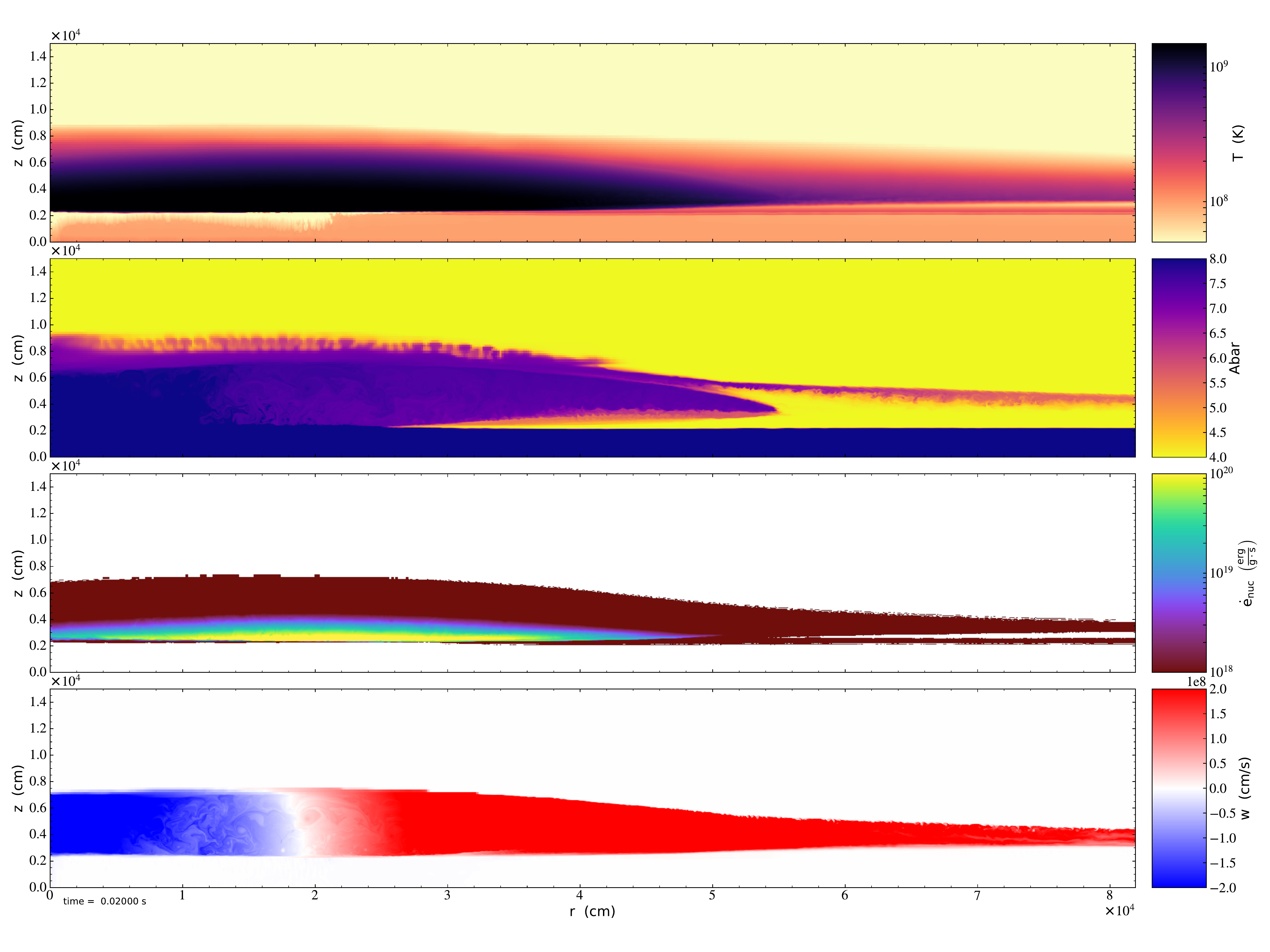}
\caption{\label{fig:10_10_lowres} Temperature, mean molecular weight,
  energy generation rate, and out-of-plane velocity for the 10/10 low
  resolution simulation at 20~ms.}
\end{figure}

Our default resolution puts $\sim 80$ zones vertically in the
``cool''  model atmosphere height.  To
understand the effects of resolution, we also performed a run with one
fewer refinement level, giving a $20~\mbox{cm}$ resolution overall (this is our
10/10-lowres run).  Figure~\ref{fig:10_10_lowres} shows the fields for this run at
20~ms.  The structure is largely the same as the 10/10 run, with
largely the same flame shape and position, and the same structure in
the energy generation rate.  Since we do not have a jump in refinement
right below the atmosphere, we don't see the temperature feature from
the initial model mapping there, but we do see some cooling in the
\isot{Ni}{56} region.  This is likely a resolution effect, again due
to the strong gradient in temperature at the base of the atmosphere.
The strong agreement in the low resolution run to the 10/10 run gives
us confidence that we are capturing the flame physics properly.

\subsection{Effects of our Approximations}

\begin{figure}[t]
\centering
\plotone{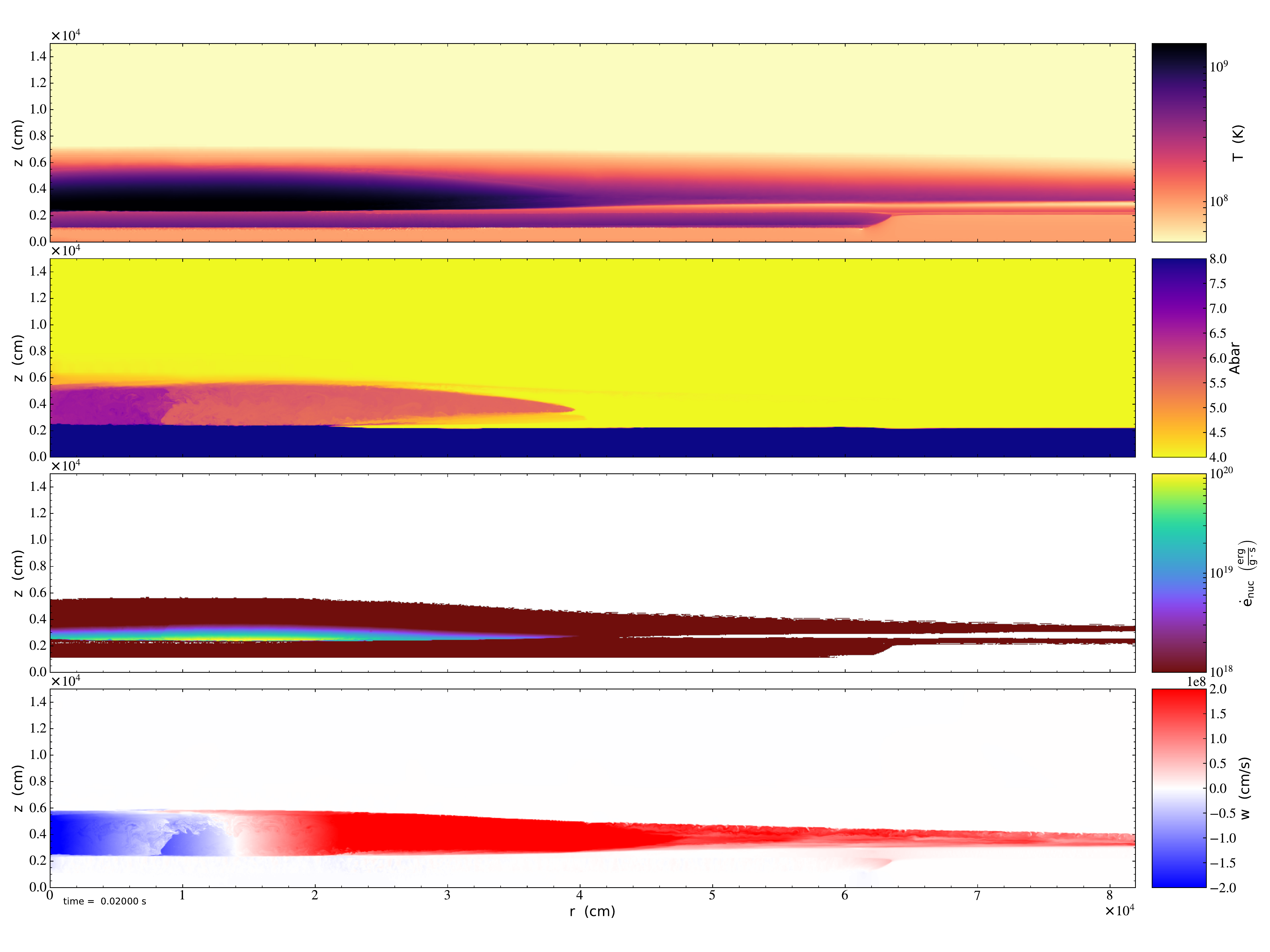}
\caption{\label{fig:5_5_overview} Temperature, mean molecular weight, energy generation rate, and out-of-plane velocity for the 5/5 simulation at 20~ms.}
\end{figure}

The results above all used a boosting of 10/10.  To see how the
results are sensitive to this boosting, we ran a simulation with a reduced
boosting of 5/5.  This is shown in Figure~\ref{fig:5_5_overview}.  The
results look qualitatively the same---a laterally propagating flame
develops that is lifted off of the bottom of the fuel layer.  The
flame has not moved out as far as in the 10/10 simulation, simply
because there is less energy release, but we expect that if we were to
run this out twice as long, the flame would have advanced to the
position seen in our 10/10 runs.  The ash is also not as evolved to as high of an $\bar{A}$.  The good agreement in the structure
of the flame seen with the lower boosting gives us confidence that the
overall aspects of the flame structure and acceleration we are seeing
are robust to the approximations we make.

\begin{figure}[t]
\centering
\plotone{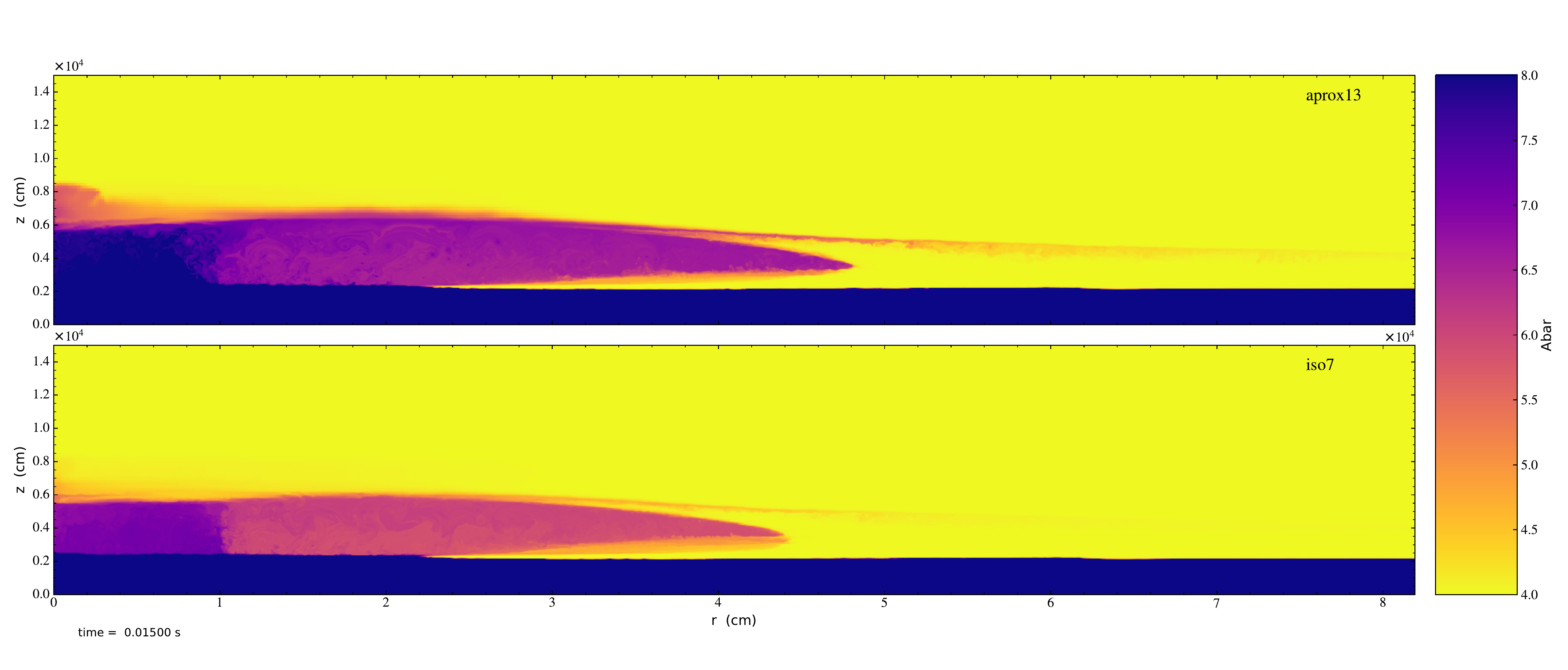}
\caption{\label{fig:network} Mean molecular weight at 15~ms comparing a run with
  {\tt aprox13} to a run with the {\tt iso7} network.}
\end{figure}

We also considered the effect of the network size.
Figure~\ref{fig:network} shows a comparison of the 10/10 boosting run
with the standard 13-isotope {\tt aprox13} network and the reduced
7-isotope {\tt iso7} network.  We see that the flame in the {\tt aprox13}
case is slightly more advanced and has a higher $\bar{A}$ than the {\tt iso7}
case.  We see in the next section that these two networks give largely
the same flame speed.  The reduced network size saves a lot of memory,
which will be useful when we transition to 3D simulations.

\begin{figure}[t]
\centering
\plotone{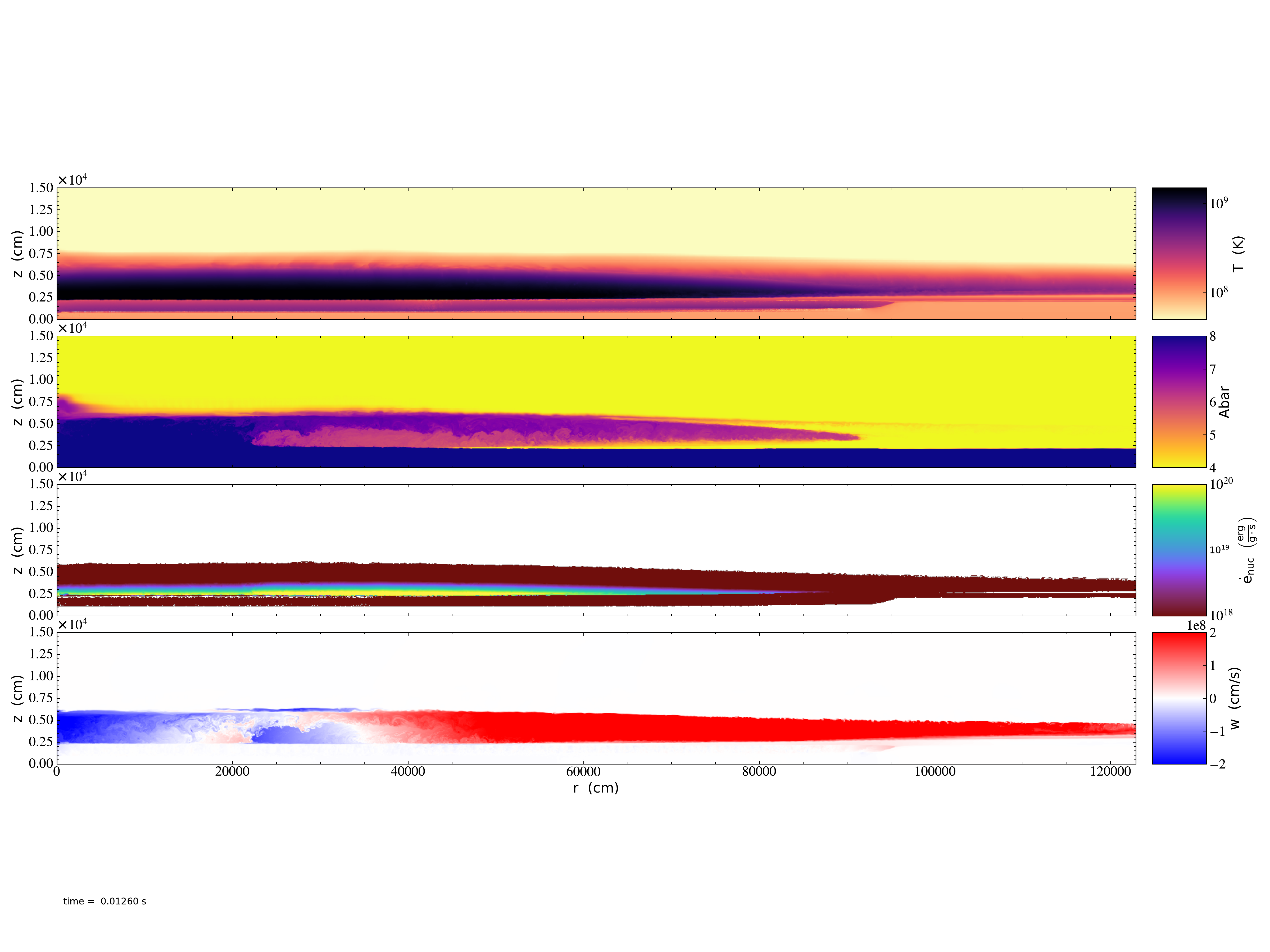}
\caption{\label{fig:10_10_slow} Temperature, mean molecular weight, energy generation rate, and out-of-plane velocity for the 10/10-1000 Hz simulation at 12.6~ms.}
\end{figure}

The final approximation to explore is our choice of rotation rate.  We ran the 10/10-1000~Hz
simulation for 12.6~ms (shorter than the 20~ms we
used for the other runs).  We also used a larger domain, $1.8432\times
10^5~\mathrm{cm} \times 3.072\times 10^4~\mathrm{cm}$ to account for
the larger Rossby radius.  Figure~\ref{fig:10_10_slow} shows the flame
structure.  Overall it looks much like the faster rotator.  In the
next section, we explore the effect of rotation on the flame
acceleration.

\subsection{Flame Propagation}

\begin{figure}[t]
\centering
\plotone{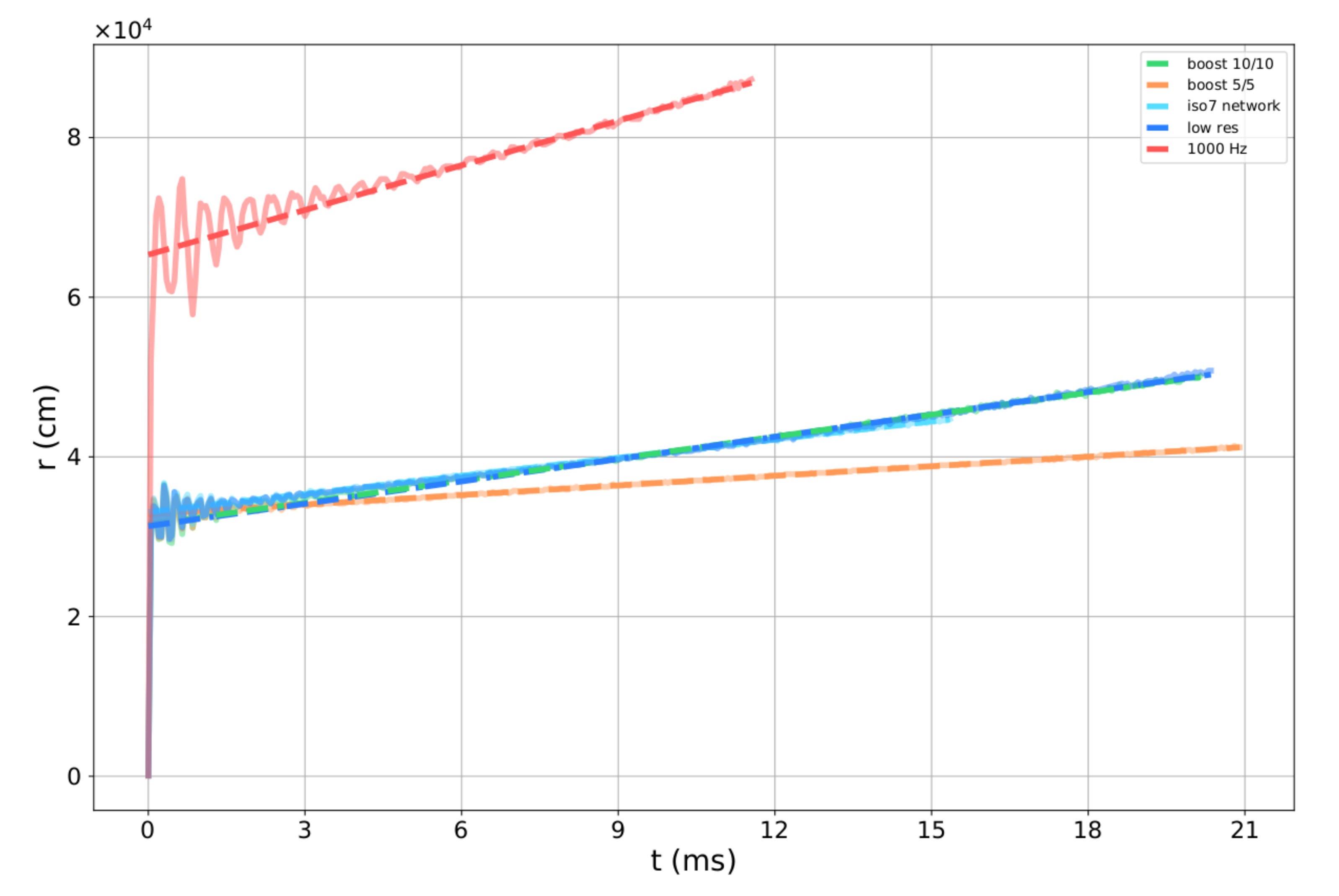}
\caption{\label{fig:flame_speed}The position of the burning front for each simulation run as a function of time. The dashed lines show linear least squares fits for $t \gtrsim 6$ ms.}
\end{figure}

To measure the propagation rate of the burning front, we first collapse our nuclear
energy generation rate ($\dot{e}_\mathrm{nuc}$) data at each time into a 1D radial
profile by averaging over the vertical coordinate. We then take the peak
$\dot{e}_\mathrm{nuc}$ value across all profiles to provide a fixed reference point.
We define the position of the flame front to be the location ahead of hottest part of
the flame where $\dot{e}_\mathrm{nuc}$ first drops to $< 0.1 \%$ of the global maximum. 
This corresponds roughly to the leading edge of the burning region. Averaging
over the vertical coordinate helps to reduce sensitivity to localized fluid motions, as does
tracking the $0.1 \%$ contour rather than a local maximum. As we see in Figure
\ref{fig:flame_speed}, the flame settles into a state of steady propagation after an initial
transient period spanning $\sim 3$ ms. The position data here are well fitted by a linear
function of time, and the resulting slope gives the velocity of the flame front.

\begin{deluxetable}{lc}
	\tablecaption{\label{table:flame_speeds_multid} Flame speeds measured in 2D calculations.}
	\tablehead{\colhead{run} & \colhead{$s_\mathrm{front}$ (km s$^{-1}$)}} 
	\startdata
	10/10 & $9.18 \pm 0.03$ \\
	5/5 & $4.00 \pm 0.01$ \\
	10/10-iso7 & $7.56 \pm 0.02$ \\
	10/10-lowres & $9.33 \pm 0.04$ \\
	10/10-1000 Hz & $18.6 \pm 0.16$ \\
	\enddata
\end{deluxetable}

Table \ref{table:flame_speeds_multid} gives the flame speed measured in each multidimensional
simulation. The 2D flames propagate at speeds about an order of magnitude faster than their 1D
counterparts (Table \ref{table:flame_speeds_1d}). The increase in flame speed is likely a product
of the larger flame surface area and hydrodynamical effects such as turbulence, wrinkling, and
convective cycles, which bring cooler fuel from ahead of the front into the hottest part of the
burning region.

All of the various approximations have the expected effects. The flame in the 5/5
boost run propagates at about half the speed of the 10/10 flame. This is consistent
with the 1D laminar flames, and is the behavior predicted by Eq.~\ref{eq:flame_scaling}. The
1000~Hz run goes 2 times faster than the 10/10 2000~Hz run, as expected from the
inverse relation between rotation rate and the ratio of burning front area to scale height \citep{cavecchi:2013}.
This confirms that the role of the Coriolis
force is to limit the rate of flame spreading by the anticipated geometrical/hydrodynamical
effect ($s_\mathrm{front} = s_L * L_R / H$). Reducing the resolution had minimal impact on
the flame speed -- the low resolution line is right on top of the fit for the standard run
in Figure \ref{fig:flame_speed}, with their slopes differing by only a few percent. Using a
smaller network also produces similar behavior to the standard run, although there is a small
reduction in speed owing to less energetic burning.


\subsection{Entrainment, Flow Features}

\begin{figure}[t]
\centering
\plotone{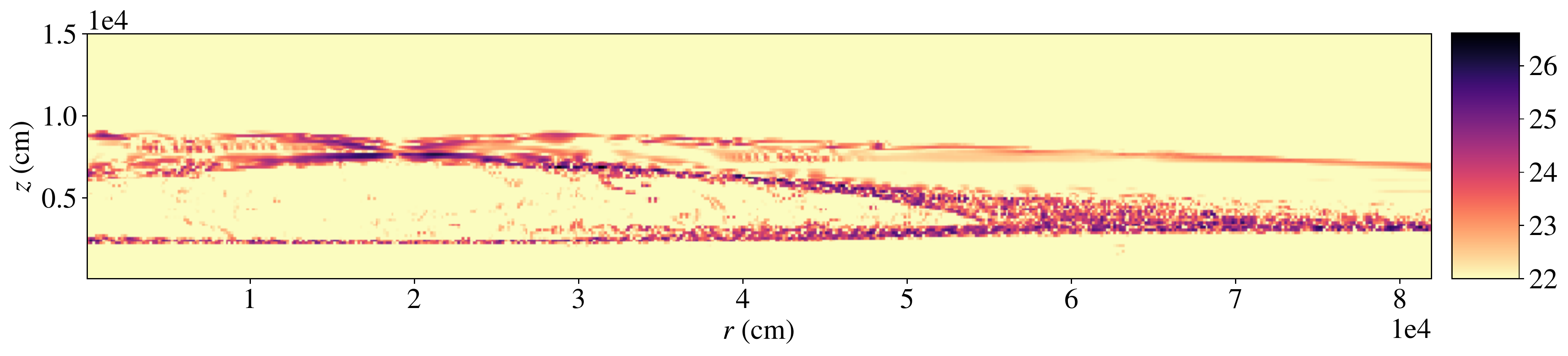}
\caption{\label{fig:baroclinicity} Baroclinicity. This plot shows
$\ln \left(\mathbf{\psi}\right)$ at time $t = 0.02~\mathrm{s}$ for the 10/10 simulation.}
\end{figure}

We explore the baroclinic instability by looking at the magnitude of the 
baroclinicity, calculated as
\begin{equation}
    \boldsymbol{\psi} = \frac{1}{\rho^2} \nablab p \times \nablab \rho,
\end{equation}
and shows the misalignment of the local density and pressure gradients
(we also explored this in \citep{Malone2014a}). As we are considering a 2D system, the
component of the baroclinicity we consider here (out of the plane)
reflects the misalignment of the fields in the plane of the
simulation. Figure~\ref{fig:baroclinicity} shows that the
baroclinicity peaks along the flame front. This baroclinicity
generates vorticity, which in turn entrains material along the surface
of the flame (see also \citealt{cavecchi:2013}). In 3D, this same vorticity
should perturb the flame front and further increase the flame speed
\citep{Cavecchi2019}.

\begin{figure}[t]
	\centering
	\plotone{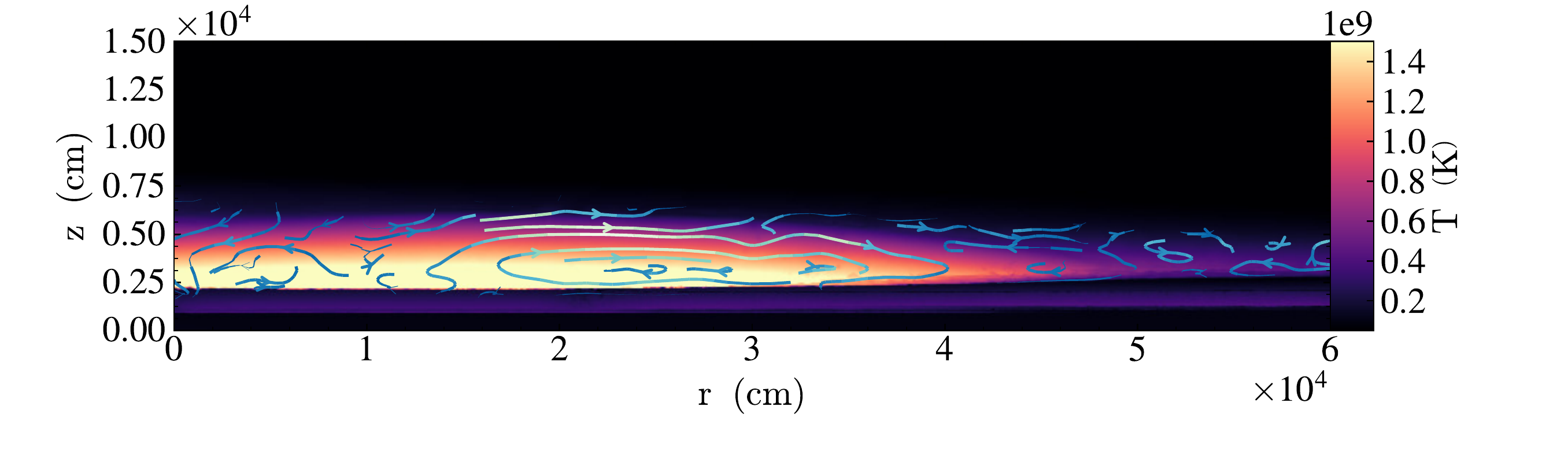}
	\caption{\label{fig:streamlines} Streamlines showing velocity field at time $t = 0.01~\mathrm{s}$ for the 10/10 simulation.}
\end{figure}

Vortical motions are also evident in a direct visualization of the velocity field, as seen in
Figure \ref{fig:streamlines}. We observe turbulence in the wake of the flame and in the
unburnt fuel ahead of the front, while a convective cycle sets up near the hottest part of the
burning region. The cycle draws the cooler fluid near the interface into the center of the flame
and drives hot ash out towards the instability at its surface, helping to facilitate the flame
spreading. Convective mixing should still be important in 3D, but we would expect much more
complicated flow patterns, with a greater contribution from small-scale features \citep{xrb3d}.

\begin{figure}[t]
\centering
\plottwo{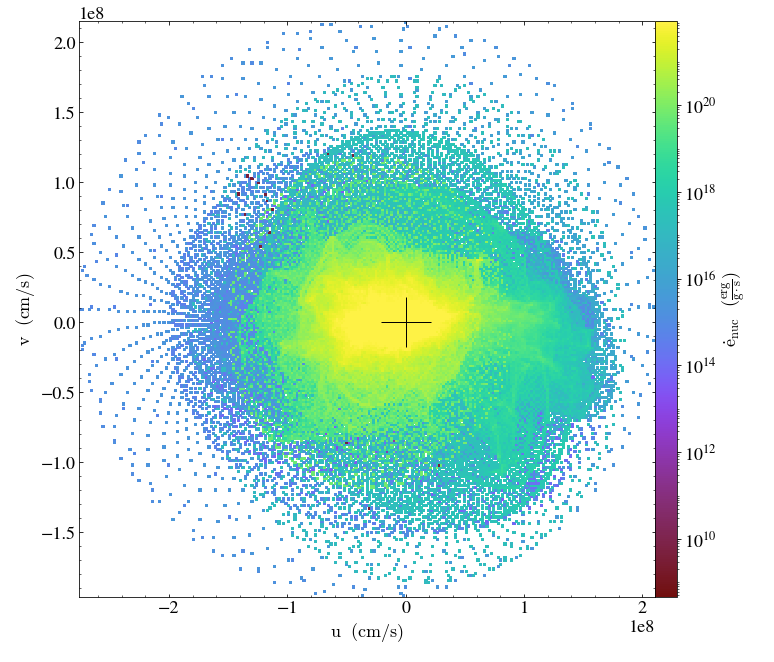}{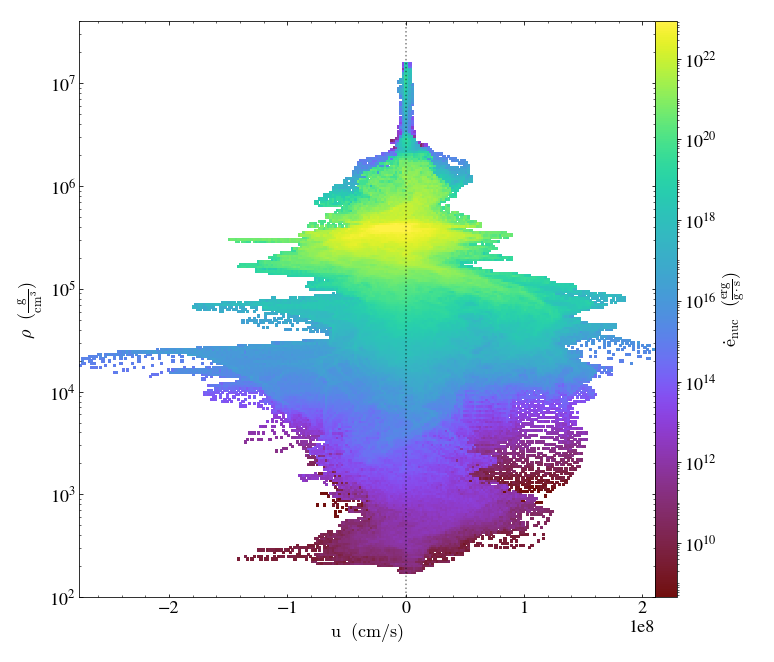}
\caption{\label{fig:phase_plots} Phase plots at time $t = 0.02~\mathrm{s}$.
\emph{Left}: Phase plot showing the energy generation rate as a function of
the $x$- and $y$-velocities. The black cross shows the location of $u = v = 0$. \emph{Right}: phase plot showing the energy
generation rate as a function of the $x$-velocity and the density.}
\end{figure}

To further demonstrate the relationships between different properties
of the flame, in Figure~\ref{fig:phase_plots} we present some phase
plots of the energy generation rate. The phase plot of the energy
generation rate as a function of the $x$- and $y$-velocities shows
that energy is preferentially generated in regions with negative
$x$-velocity. This most likely corresponds to the burning along the
underside of the flame front, where fuel has been entrained along the
surface of the flame and so is moving in the opposite direction to the
direction the flame is propagating in. This is further supported by
the phase plot of the energy generation rate as a function of the
$x$-velocity and the density, which shows that the energy generation
rate peaks at $\rho \sim 3 \times 10^5~\mathrm{g/cm}^3$, at the base
of the flame.  The energy generation rate also depends strongly on density.

In the first plot, there are ``loops" of
points of similar $\dot{e}_\mathrm{nuc}$ in the outer edges of the plot.
These are likely to correspond to the vortices that appear within the flame,
where the fluid moves in a circular motion in $u-v$ phase space.

\begin{figure}[t]
\centering
\plotone{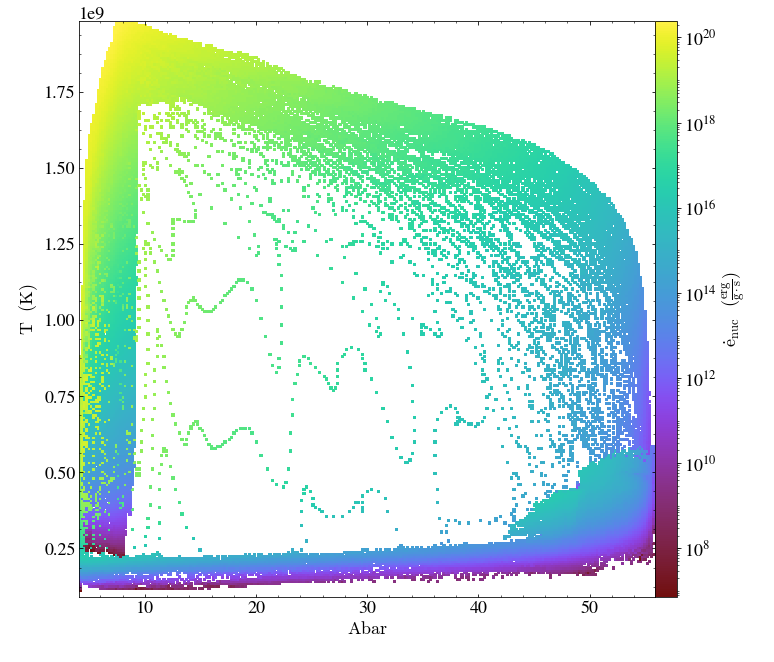}
\caption{\label{fig:abar_temp} Phase plot of energy generation rate as a function of $\bar{A}$ and temperature at time $t = 0.02~\mathrm{s}$.}
\end{figure}

Figure~\ref{fig:abar_temp} shows the energy generation rate as a function of the
temperature and $\bar{A}$. 
The energy generation rate peaks at low $\bar{A}$, where there is a high fraction
of unburnt material. The burning raises the temperature of this material, such
that the peak temperature coincides with the peak energy generation rate. The
material then cools again as the burning converts the fuel into ashes,
increasing $\bar{A}$ and reducing the energy generation rate as there is
less available fuel. Along the base of the plot we see cool unburnt fluid and
ashes. In the center of the plot there are thin `trails' in phase space, which
could correspond to less common reaction pathways in the reaction network.

\section{Discussion}

We have shown the results of our fully hydrodynamical,
multidimensional simulations of flame propagation through the
atmosphere of a neutron star. By using the fully compressible
hydrodynamics equations, we are able to capture the vertical dynamics
of the system. To accurately model the flame propagation, it is
necessary to sufficiently resolve the scale height of the
atmosphere. It is also important that there is a thin transition
between the underlying neutron star and the atmosphere, to ensure that
the peak temperature occurs at the correct base density.

In order to satisfy these requirements whilst allowing the simulations
to be computationally feasible, we used several approximations: a
higher than normal rotation rate, a boosted flame, a simplified
reaction network, and a 2D axisymmetric model for the flow. The first two of
these were used to reduce the model's spatial and temporal
scales. Using the simulation framework developed here, these both can
be relaxed in the future, at the cost of more computer time. The same
goes for 2D axisymmetry vs.\ full 3D---the only difference is computer time, and
our future calculations will explore the 3D evolution and compare to
the 2D simulations presented here. In particular, in 3D we will be
able to explore shear instabilities at the flame front.  We can also
capture the baroclinic instability \citep{Cavecchi2019} and the
competition between it and shear.  Larger networks are a
straightforward change, and already supported in \castro\ using the
\pynucastro\ framework~\citep{pynucastro} and JINA ReacLib rate
database~\citep{reaclib}.

In addition to extending our models to 3D and relaxing the
approximations used in this study, in the future we plan to perform a
number of additional further studies.  These include investigating the
effects of ignition latitude and different initial models. We also
plan to model mixed H/He bursts. This will require a different
reaction network, and perhaps different resolution
requirements. A long term goal is to include magnetohydrodynamics in
our models. Although the magnetic fields of neutron stars exhibiting
X-ray bursts are relatively weak, with $B \lesssim 10^8 - 10^9$G
\citep{mukherjee2015magnetic} (at least compared to e.g. magnetars,
which have $B\lesssim 10^{15}$G), it has been found by
\citet{art-2016-cavecchi-etal} that even weak magnetic fields could
have a non-negligible effect on the flame propagation, reducing
confinement due to the Coriolis force and leading to increased flame
speeds. Ultimately, we wish to link our simulations to observed light
curves. To do this, we will need to explore radiation transport in
order to model how the burst energy propagates through the outer
layers of the neutron star atmosphere.








In addition to exploring other aspects of the XRB physics, there are several changes to the algorithm used to
model these XRBs we will pursue.  First, as shown in \citet{castro-sdc}, we have
developed a fourth-order (in space and time) method for coupling
hydrodynamics and reactions that should greatly improve the accuracy
of the simulations.  We expect that by using this new high-order
algorithm we can drop a level of refinement from the simulations while
still accurately modeling the evolution.  We have also ported
\castro\ to GPUs, giving an order of magnitude performance boost on
nodes with both CPUs and GPUs.  Flames without any boosting are currently running
using the new GPU-enabled \castro and will be the focus of the next study.


\acknowledgements \castro\ is open-source and freely available at
\url{http://github.com/AMReX-Astro/Castro}.  The problem setups used
here are available in the git repo as {\tt flame} and {\tt
  flame\_wave}.  The work at Stony Brook was supported by DOE/Office
of Nuclear Physics grant DE-FG02-87ER40317 and the SciDAC program DOE grant DE-SC0017955.  MZ acknowledges support
from the Simons Foundation.  YC was supported by the
European Union Horizon 2020 research and innovation program under
the Marie Sklodowska-Curie Global Fellowship grant agreement No.\
703916.  This research used resources of the National Energy Research
Scientific Computing Center, a DOE Office of Science User Facility
supported by the Office of Science of the U.~S.\ Department of Energy
under Contract No.\ DE-AC02-05CH11231.  This research used resources
of the Oak Ridge Leadership Computing Facility at the Oak Ridge
National Laboratory, which is supported by the Office of Science of
the U.S. Department of Energy under Contract No. DE-AC05-00OR22725,
awarded through the DOE INCITE program.  This research has made use of
NASA's Astrophysics Data System Bibliographic Services.

\facilities{NERSC, OLCF}

\software{AMReX \citep{amrex_joss},
          Castro \citep{castro},
          GCC (\url{https://gcc.gnu.org/}),
          linux (\url{https://www.kernel.org/}),
          matplotlib (\citealt{Hunter:2007}, \url{http://matplotlib.org/}),
          NumPy \citep{numpy,numpy2},
          python (\url{https://www.python.org/}),
          valgrind \citep{valgrind},
          VODE \citep{vode},
          yt \citep{yt}}

\appendix

\section{Diffusion Tests}
\label{app:diffusion}

To test our implementation of diffusion, we created a unit test in
\castro\ that just uses the diffusion solver.  This test uses the
standard Gaussian initial conditions---the diffusion of a Gaussian
remains Gaussian, with a lower amplitude and wider width.  This is run
with a constant diffusion coefficient. The point of this test is to
demonstrate that the implementation done in a predictor-corrector
form is second-order accurate.

We use a 2D axisymmetric geometry, with a reflection boundary on the symmetry axis (left)
and Neumann boundaries on all other sides.  In this geometry, the solution behaves like
a spherical problem, and has the form:
\begin{equation}
  T(r,t) = T_1 + (T_2 - T_1) \left (\frac{t_0}{t + t_0} \right )^{3/2} e^{-r^2/(4 D (t + t_0))}
\end{equation}
where $t_0$ is a small time used to set the initial width of the
Gaussian, $r$ is the distance from the origin, and $D$ is the
diffusion coefficient (see, e.g., \citet{SWE_MYRA09}).  $T_1$ is the ambient temperature and $T_2$ is
the temperature of the peak of the Gaussian.  We use $D = 1$, $T_1 = 1$, $T_2 = 2$,
and $t_0 = 0.001$.

\begin{deluxetable}{rLc}
	\tablecaption{\label{table:diffusion} Convergence of diffusion test problem.}
	\tablehead{\colhead{base resolution} & \colhead{L-inf error} & \colhead{convergence rate\tablenotemark{a}}}
	\startdata
\cutinhead{no refinement}
	32  & 5.995\times 10^{-3} & \multirow{2}{*}{2.04} \\
        64  & 1.477\times 10^{-3} & \multirow{2}{*}{2.01} \\
       128  & 3.664\times 10^{-4} & \multirow{2}{*}{2.00} \\
       256  & 9.139\times 10^{-5} & \\
\cutinhead{one level of refinement}
        32  & 1.480\times 10^{-3} & \multirow{2}{*}{2.01} \\
        64  & 3.685\times 10^{-4} & \multirow{2}{*}{1.97} \\
       128  & 9.406\times 10^{-5} & \multirow{2}{*}{1.94} \\
       256  & 2.446\times 10^{-5} & \\
	\enddata
\tablenotetext{a}{Convergence, $n$, is defined between 2 successive resolutions as $n = \log_2(e_{2h}/e_h)$,
 where $e_{2h}$ is the error for the coarser resolution run and $e_h$ is the error for the finer resolution run.}
\end{deluxetable}

\begin{figure}[t]
\centering
\epsscale{0.5}
\plotone{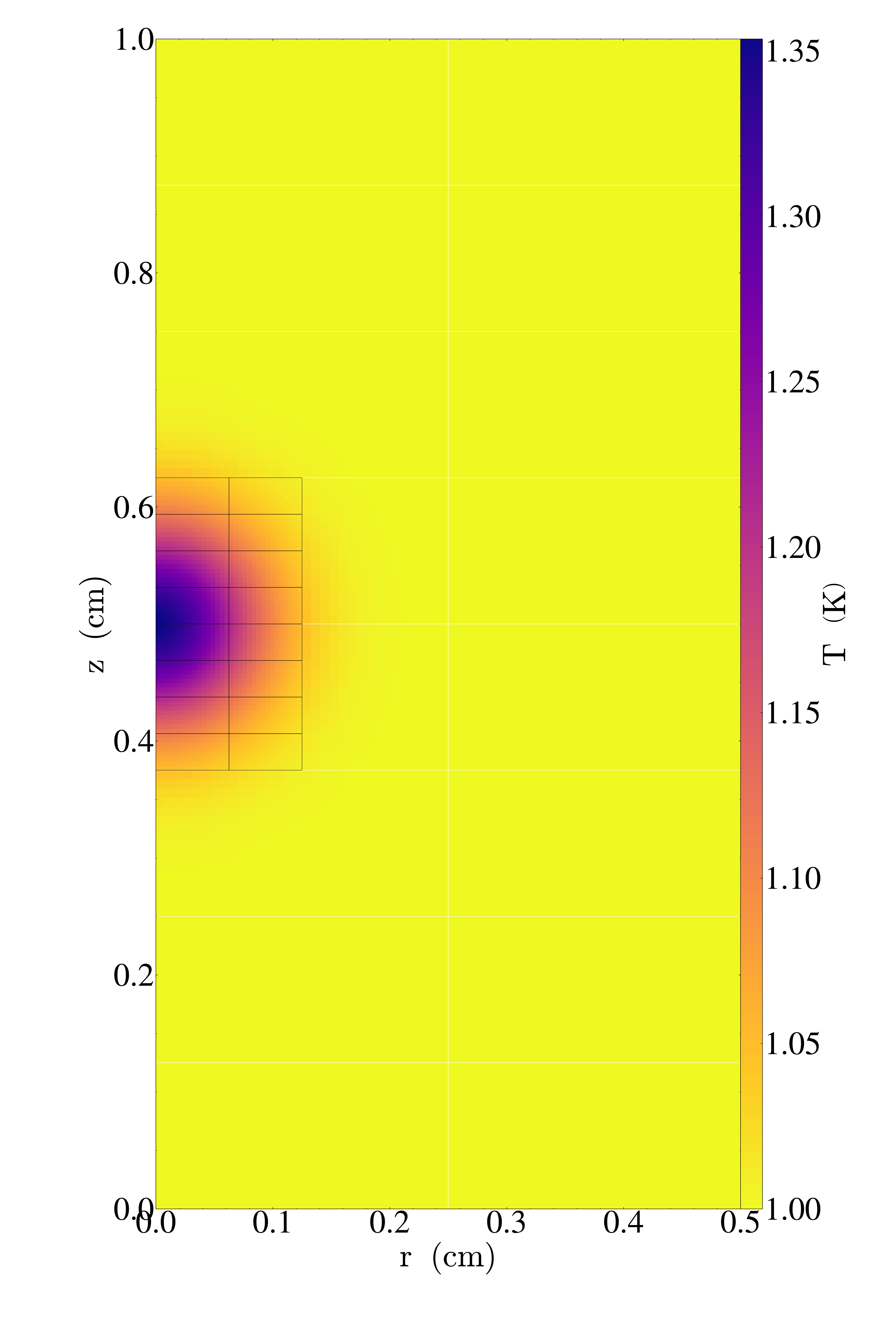}
\caption{\label{fig:diffusion} Temperature for the diffusion test problem at $10^{-3}~\mbox{s}$
for a diffusion test with a $64\times 128$ zone base grid and one level of refinement.}
\epsscale{1.0}
\end{figure}

Table~\ref{table:diffusion} shows the L-inf error against the analytic
solution for the diffusion test problem.  The tests were run in two
fashions: no AMR, and AMR with one level of refinement (a jump of
$2\times$).  In both cases we see second-order convergence of the
error.  Figure~\ref{fig:diffusion} shows the temperature at the end of
the simulation of the $64\times 128$ base grid plus one level of
refinement simulation, with the grids shown to illustrate where the
refinement jump is.

\bibliographystyle{aasjournal}
\bibliography{ws}

\end{document}